\documentclass[fleqn,usenatbib]{mnras}

\usepackage{newtxtext,newtxmath}
 
\usepackage[T1]{fontenc}

\DeclareRobustCommand{\VAN}[3]{#2}
\let\VANthebibliography\thebibliography
\def\thebibliography{\DeclareRobustCommand{\VAN}[3]{##3}\VANthebibliography}

\usepackage{graphicx}	
\usepackage{amsmath}	
\usepackage{amssymb}	
\usepackage[separate-uncertainty=true]{siunitx}
\usepackage{multirow}
\usepackage{color}
\usepackage{csquotes}
\usepackage[normalem]{ulem}

\newcommand{\id}[1]{_\textrm{#1}}

\newcommand{\Ms}{M_{\sun}}
\newcommand{\M}[1]{\SI{#1}{\Ms}}

\newcommand{\hmpc}{h^{-1}\,\mathrm{Mpc}}

\DeclareSIUnit{\pc}{pc}
\DeclareSIUnit{\year}{yr}
\DeclareSIUnit{\particles}{particles}  
\DeclareSIUnit{\keV}{\kilo\electronvolt}
\DeclareSIUnit{\kmsMpc}{\kilo\meter\per\second\per\mega\pc}
\DeclareSIUnit{\Myr}{\mega\year}
\DeclareSIUnit{\kpc}{\kilo\pc}
\DeclareSIUnit{\Mpc}{\mega\pc}
\DeclareSIUnit{\H}{H}
\DeclareSIUnit{\erg}{erg}
\DeclareSIUnit{\pcm}{\per\centi\meter\cubed}
\DeclareSIUnit{\Msun}{M_{\sun}}
\DeclareSIUnit{\Zsun}{Z_{\sun}}
\DeclareSIUnit {\h}{\ensuremath{\mathit{h}}}

\newcommand{\rom}[1]{\uppercase\expandafter{\romannumeral #1\relax}}

\defcitealias{Moster2013}{M13}
\defcitealias{Behroozi2013}{B13}
\defcitealias{Behroozi2019}{B19}

\title[Mechanical feedback from stellar winds]{Mechanical feedback from stellar winds with an application to galaxy formation at high redshift}

\author[]{
Yvonne A. Fichtner$^{1}$\thanks{E-mail: yfichtner@astro.uni-bonn.de}, 
Luca Grassitelli$^{1}$, 
Emilio Romano-D\'{\i}az$^{1}$ 
and Cristiano Porciani$^{1}$
\\
$^{1}$ Argelander-Institut für Astronomie, Auf dem Hügel 71, D-53121 Bonn, Germany\\
}

\date{Accepted 2022 March 17. Received 2022 March 17; in original form 2022 January 15}

\pubyear{2022}

\begin{document}
\label{firstpage}
\pagerange{\pageref{firstpage}--\pageref{lastpage}}

\maketitle

\begin{abstract}
We compute different sets of stellar evolutionary tracks in order to quantify the energy, mass, and metals yielded by massive main-sequence and post-main-sequence winds. Our aim is to investigate the impact of binary systems and of a metallicity-dependent distribution of initial rotational velocities on the feedback by stellar winds. We find significant changes compared to the commonly used non-rotating, single-star scenario. The largest differences are noticeable at low metallicity, where the mechanical-energy budget is substantially increased.
So as to establish the maximal (i.e. obtained by neglecting dissipation in the near circumstellar environment) influence of winds on the early stages of galaxy formation,
we use our new feedback estimates to simulate the formation and evolution of a sub-$L_*$ galaxy at redshift 3 (hosted by a dark-matter halo with a mass of $1.8\times 10^{11}$\,M$_\odot$) and compare the outcome with simulations in which only supernova (SN) feedback is considered.
Accounting for the continuous energy injection by winds
reduces the total stellar mass, the metal content, and the burstiness of the star-formation rate as well as of the outflowing gas mass. However, our numerical experiment suggests that the enhanced mechanical feedback from the winds of rotating and binary stars has a limited impact on the most relevant galactic properties compared to the non-rotating single-star scenario. Eventually, we look at the relative abundance between the metals entrained in winds and those ejected by SNe and find that it stays nearly
constant within the simulated galaxy and its surrounding halo.
\end{abstract}




\section{Introduction}

Modern astrophysics 
relies on the inclusion of stellar-feedback processes
in order to explain the low star-formation efficiency and baryon fraction observed in galaxies together with the metal enrichment of the circumgalactic and intergalactic media.
The most commonly considered form of stellar feedback is the injection of mass, heavy elements, energy, and momentum by supernova (SN) explosions
\citep[e.g.][]{Dekel1986}.
The energy budget per unit mass of stars formed is dominated by core-collapse SNe \citep[see e.g. fig. 10 in the review by][]{Benson2010PhR}.
Each SN deposits 1-10 M$_\odot$ of stellar ejecta initially moving at $\sim10^4$ km s$^{-1}$ (much larger than the sound speed in the surrounding medium, thus leading to a blast wave) with a total kinetic energy of $\sim10^{51}$ erg. The time evolution of the supernova remnant (SNR) produces a final momentum of a few $\times\, 10^5$ M$_\odot$ km s$^{-1}$, slightly depending on the properties of the environment \citep[e.g.][]{KimOstriker2015}. The characteristic length scale of this phenomenon 
(say the radius of the dense shell that forms at the outer edge of the SNR between the Sedov-Taylor and the pressure-driven snowplow stages) ranges between a few and a few tens 
of pc determined by the ambient density. Multiple events clustered in space and time can build up superbubbles of hot gas in the interstellar medium (ISM) that break through the galactic disc and vent material into the halo \citep[][]{Norman1989} or even beyond.
It is widely believed that SNe play a key role in the self-regulation of star formation within galaxies: gravity and cooling cause the gas to reach high densities and turn into stars thus producing feedback which drives some gas back to lower densities. 
Finally, SN explosions are thought to govern the small-scale structure of 
the ISM which comprises multiple `thermal phases' in approximate pressure balance with one another.
The ISM is often modelled as consisting
of cold clouds (which dominate the mass budget) embedded in a hotter volume-filling inter-cloud medium.
The physical conditions of the multi-phase ISM are determined by
mass exchanges between the phases and  the energy injection due to SNe \citep[][]{McKee1977}. In particular, the hot phase cools down slowly
through the evaporation of the cold clouds.

Several 
lines of reasoning suggest that additional forms of stellar feedback should be included
in galaxy-formation models
\citep[e.g.][]{Hopkins2011, Brook2012, Stinson2012}
although it is challenging to differentiate this need from the limitations of the sub-grid models for SN feedback and star formation
\citep{Rosdahl2017, Smith2019}.
Beyond SN explosions, there are several physical phenomena through which stellar feedback could affect the ISM, namely: proto-stellar jets, radiation pressure, photoionization, photo-electric heating, stellar winds 
and
cosmic-ray acceleration at SNRs.
These processes take place
on rather small scales (ranging from a fraction of a pc to 10 pc)
and can in principle play an important role in dispersing the gas after the onset of star formation.
In particular, pre-SN or `early'
feedback is thought to strengthen the impact of the succeeding SN blast wave \citep[e.g. ][]{Agertz2013, Stinson2013, Geen2015, Hopkins2018} by removing gas from the birth giant molecular cloud (GMC). 
This scenario is supported by observations showing that gas is cleared out of the star-formation site before the first SN explosion \citep{Barnes2020} and that GMCs are dispersed within \SI{3}{\Myr} after the emergence of unembedded high-mass stars \citep{Chevance2020}. 

Due to computational cost, 
high-resolution simulations
of stellar feedback within the ISM often concentrate on only one or two processes, namely individual and clustered SNe
\citep{Gatto2015, IffrigHennebelle2015, KimOstriker2015, WalchNaab2015, Girichidis2016, Kortgen2016},
SNe and photoionization \citep{Geen2016},
SNe and stellar winds \citep{Rogers2013, Gatto2017},
photoionization \citep{DaleBonnell2011, Haid2019, Bending2020},
photoionization and radiation pressure \citep{Howard2016, Kim2016, Ali2018, Kim2018},
photoionization and winds \citep{Dale2014, Haid2018},
radiation pressure \citep{SkinnerOstriker2015},
winds and radiation pressure \citep{SilichTenorioTagle2013}.
However, some recent simulations account for stellar winds,
photoionization and SNe \citep{Geen2015, Lucas2020, Rathjen2021}.
\citet{Stinson2013} developed an approximate numerical scheme to account for the collective effect of
early stellar feedback in
low-resolution simulations of galaxy formation.
In this case,
10 per cent of the UV luminosity of a stellar population 
is injected as thermal
energy before any SN events take place. 
Consequently, the gas immediately heats up to temperatures
$T > 10^6$~K and then 
rapidly cools down to $10^4$ K thus creating a lower density medium than in the absence of early feedback.
This broadly mimics the formation of an \ion{H}{ii} region and effectively prevents star formation in the regions immediately surrounding young stellar clusters.
On the other hand,
\citet{Hopkins2011} developed a kinetic-feedback
scheme to deposit the momentum imparted by radiation, SNe, and stellar winds
in higher-resolution simulations.
This is a sub-grid model which aims
to describe physics taking place
within GMCs and stellar clusters.
This effort eventually led to the
Feedback In Realistic Environments (FIRE) project \citep{Hopkins2014} and
its updated version FIRE-2 \citep{Hopkins2018}. 
In parallel, 
\citet[][see also \citealt{AgertzKravtsov2015}]{Agertz2013} presented a sub-grid model for stellar feedback which takes into account the time-dependent injection of energy, momentum, mass and heavy elements
from radiation pressure, stellar winds, SNe type
II and Ia into the surrounding ISM.
Undertaking a similar endeavour, \citet{Marinacci2019}
presented the
Stars and MUltiphase Gas in GaLaxiEs  (SMUGGLE) model which has been then 
combined with a radiative-transfer scheme
in \citet{Kannan2019}.
Recent simulations of individual galaxies based on these schemes seem to self-consistently generate prominent galactic fountains
and sustain inefficient (feedback-regulated) star formation for long time-scales, in agreement with observations
\citep[e.g.][]{Hu2016, Wetzel2016, Hu2017, LiBryan2017, Hopkins2018, Emerick2019, Lahen2019,   Wheeler2019, Agertz2020, Emerick2020, Lahen2020, LiTonnesen2020, LiLi2020,  Gutcke2021}.

In this work, we concentrate on the mechanical impact from stellar winds and neglect other forms of early feedback (e.g. radiative).
Massive stars show radiation-driven outflows
where material escapes the stellar surface with velocities
of $\sim1000$ km s$^{-1}$ and mass-loss rates of $\sim10^{-6}$ M$_\odot$ yr$^{-1}$ 
\citep[see][for a recent review]{vink2021r}. 
The energy injected into the ISM
by winds over the lifetime of a massive star can be comparable to the mechanical energy of the subsequent SN explosion.
Theoretical models and numerical simulations suggest that 
SN explosions should always take place within wind-blown cavities surrounded by dense shells (with radii $>10$ pc) that have been seeded with nuclear processed material  \citep{TenorioTagle1990, TenorioTagle1991, Rozyczka1993, SmithRosen2003, Dwarkadas2005, Dwarkadas2007, ToalaArthur2011, Geen2015}.
However, 
in stellar clusters, the fast stellar winds collide with each other and with the clumpy ISM producing shocks that heat the gas up. In this complex configuration, the wind energy can be lost via several channels \citep[e.g.][]{Rosen2014} and it is unclear whether stellar winds can drive the bulk motion of a cool, dense super shell surrounding the whole star-forming region.
Recent simulations give contrasting answers. For instance,
\citet{Rey-Raposo2017} find that winds act as an effective source of kinetic and thermal energy for the ISM while \citet{Lancaster2021} show that the bulk of the wind energy is lost due to turbulent mixing followed by radiative cooling. This loss, however, 
might be inhibited by magnetic fields
\citep{Rosen2021}. 
What is certain is that
the presence of a wind-driven bubble dramatically increases
the fraction of SN energy which
is retained by the ISM in the form of kinetic energy \citep{Rogers2013, Fierlinger2016}.
In fact, the SN
energy leaks through the chimneys (low-density regions) dug by the winds.
On the other hand, studies that investigate the impact on to the ISM of ionising radiation from massive stars 
find a limited impact from stellar winds \citep[e.g.][]{Dale2014, Ngoumou2015, Geen2021}.
An important caveat worth mentioning here is that the various studies cited above adopt different approximations to model the small-scale structure of the ISM and therefore cannot be always directly compared.

Most simulations of galaxy formation
\citep[e.g. ][]{Agertz2013, Hopkins2018}
rely on \textsc{Starburst99} \citep{Leitherer1999, Vazquez2005, Leitherer2010, Leitherer2014}
to derive the mechanical input of winds
from single stars.
In order to bracket the distribution of
rotation velocities in stellar populations, the latest version of
\textsc{Starburst99} includes the Geneva 2012/13 stellar models with two rotation velocities (either zero or 40 per cent of the break-up velocity on the zero-age main-sequence) and two metallicity values.
However, this is a rather limited set of evolutionary models covering a limited parameter range, which might not be fully representative of the rotational velocity distribution of stars and its metallicity-dependence.

We use the state-of-the-art open-source stellar-evolution code
Modules for Experiments in Stellar Astrophysics \citep[\textsc{Mesa},][]{Paxton2011,Paxton2013,Paxton2015,Paxton2018,Paxton2019}
in order to investigate a broader range
of rotational velocities and metallicities.
Moreover, since very massive stars have been observed in the Tarantula Nebula \citep[e.g.][]{Doran2013} in the Large Magellanic Cloud (LMC),
we consider stellar models with initial
masses up to nearly 160 M$_\odot$ (a factor of 1.3 larger than in \textsc{Starburst99}).
Finally, we account for the fact that 
the majority of stars are born
in binary systems \citep{Sana2012,Sana2013}.
The observational evidence for such large binary fraction
led to a paradigm change in our understanding of stellar evolution.
As one (or both) of the stars in a binary system fill their Roche lobes, a phase of mass transfer takes place. Material is exchanged between the companions through the first Lagrangian point or lost by the system. The outermost layers of the donor are stripped off and
eventually accreted on to the secondary star.
This significantly alters the masses and spectroscopic appearances of the stars 
and generates evolutionary sequences otherwise unattainable in a single-star scenario \citep[e.g.][]{demarco2017}, influencing the mass-loss and the rotation rates of stars. In some cases, the interaction and mass exchange can be unstable, leading to merger events.
Although numerous uncertainties still exist regarding the modelling of binary systems,
it is becoming increasingly clear that
more realistic estimates of stellar feedback cannot ignore the impact
of stellar multiplicity. This applies also to radiative feedback as
interacting binaries  
enhance the production of hydrogen- and helium-ionizing photons \citep[e.g.][]{Eldridge2017,gotberg2017} and harden the spectra of a stellar population \citep{gotberg2019}.
Of particular interest are stripped helium stars, i.e. massive helium stars produced by binary interaction which emit the majority of their light at wavelengths shorter than the Lyman limit \citep{Stanway2016,gotberg2019} on a time-scale beyond few Myr.

As a first sample application, we perform a series of zoom-in cosmological simulations 
following the formation of the central galaxy in a dark-matter (DM) halo of mass
$M=1.8\times 10^{11}$ M$_\odot$ at
redshift $z=3$. By considering different
feedback models, we investigate the impact of stellar winds from rotating stellar models and binary systems on the resulting galaxy.
We obtain objects with a stellar mass of a few $\times \,10^9$ M$_\odot$ in line with those routinely observed in deep optical and infrared imaging surveys \citep[see e.g. fig. 1 in][]{Grazian2015}.
Note that, in the standard cosmological model, main-progenitor haloes of the selected mass at $z=3$ end up, on average, within haloes with mass 
$M\simeq 10^{12}$ M$_\odot$ at the present time
\citep[e.g.][]{Wechsler2002}.
In this sense, our study is also relevant to Milky Way-like galaxies at $z=0$.


The paper is structured as follows. In Section~\ref{sec:wind_feedback} (and in the appendices), we introduce our suite of stellar evolutionary models and derive the mechanical and chemical yields of stellar winds as a function of metallicity. In Section~\ref{sec:sim_fdbk_over}, we briefly review the implementations of stellar feedback in simulations and describe 
our own.
The specifics of our cosmological simulations are given in Section~\ref{sec:numerical_methods} while, in Section~\ref{sec:results}, we study the properties of the simulated galaxies. Our main results and conclusions are listed in Section~\ref{sec:Summary}.



\section{Stellar evolution models} 
\label{sec:wind_feedback}
\begin{figure*}
	\centering
	\includegraphics[width=1\textwidth]{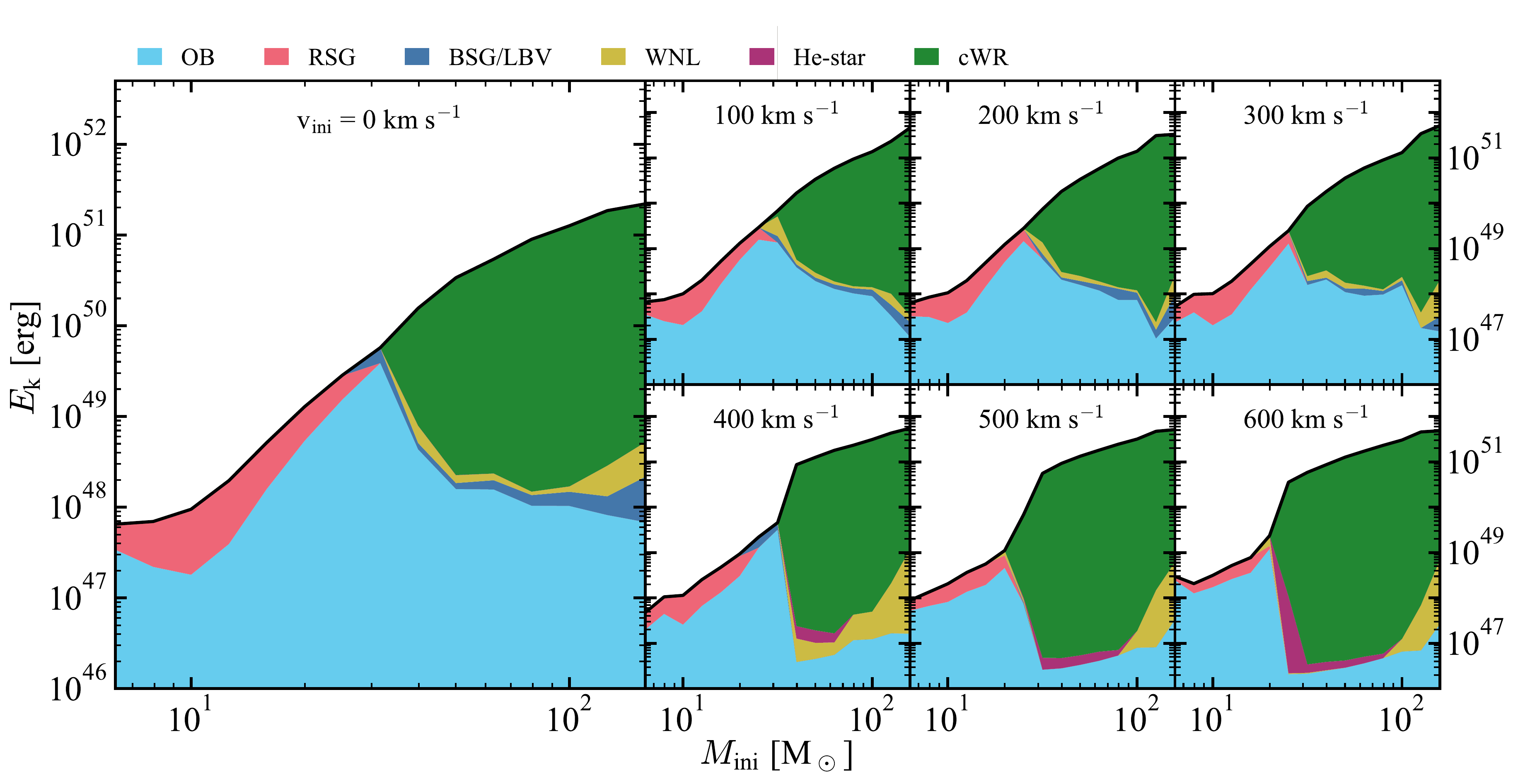}
	\caption{
	Total kinetic energy (black solid line) injected by winds into the ISM
	during the lifetime of a stellar model with initial mass $M_\mathrm{ini}$ and metallicity $Z=0.004$. Different panels refer to different rotation velocities as indicated by the labels.
	In order to distinguish the relative contributions from different evolutionary stages (see footnote~\ref{footdef}), we shade the area between the black curve and the bottom axis using different colours. The vertical extent of each colour indicates the fraction of energy released during each phase using a linear scale (e.g. a contribution of 50 per cent covers half of the distance between the bottom axis and the black curve).
	}
	\label{fig:Methods_Single_Initialmass_Energy}
\end{figure*}
We use the stellar evolution code
\textsc{Mesa} to compute different sets of models.
For the mass-loss by stellar winds, we adopt the so-called `Dutch' prescription based on
\citet{Vink2001} for OB stars, \citet{Nugis2000} for the Wolf-Rayet (WR) regime, and \citet{deJager1988} for late-type stars.  
For convection, we adopt the mixing-length theory  \citep{Bohm-Vitense1958,Henyey1965}  with a mixing-length parameter of $2$ in the framework of the MLT++ scheme. This is motivated by the numerical difficulties commonly found in inflated stellar envelopes \citep{Sanyal2015}. For the overshooting, semi-convection, and thermohaline-mixing parameters, we assume the values of $0.345$, $1$, and $1$, respectively \citep[similar to][]{Brott2011}. Any remaining numerical parameters are set as in \citet{Marchant2016}. We neglect magnetic fields.

\subsection{Wind feedback from single stars}
\label{sec:wind_feedback_single}
Our grid of
models for single stars spans a mass range of $\log(M\id{ini}/\si{\Msun})=$ \num{0.8}--\num{2.2} in steps of $\log(M\id{ini}/\si{\Msun})=0.1$, where $\log$ is a short for $\log\id{10}$. We consider eight different metallicities, namely \numlist{0.0001;0.0004;0.0007;0.001;0.002;0.004;0.008;0.02} and vary the initial surface rotation velocities, $\upsilon_\mathrm{ini}$, between $0$ and $600$~\si{\kilo\meter\per\second} in intervals of \SI{100}{\kilo\meter\per\second}. 
All our models are computed until core-helium exhaustion.

For a stellar model with mass $M$,
radius $R$, luminosity $L$, surface hydrogen fraction $X$, metallicity $Z$, mass-loss rate $\dot{M}$, surface rotation velocity $\upsilon\id{rot}$ and critical rotation velocity $\upsilon\id{crit}$ (determined by \textsc{Mesa}),
we estimate the rate of kinetic energy ejected in the form of winds using \citep{Leitherer1992,Puls2008}
\begin{equation}
\dot{E}\id{k}=\frac{1}{2}\,\dot{M} \upsilon_\infty^2
\label{eq:Stellar_energy}
\end{equation}
with the terminal wind velocity
\begin{equation}
\upsilon_\infty=d\,\sqrt{\frac{2\,G\,M}{R}\,(1-\Gamma\id{es}) }\,f\id{rot} \, \left(\frac{Z}{0.02}\right)^{0.13},
\label{eq:Stellar_terminal_wind_velocity}
\end{equation}
where $d$ is the factor relating the terminal and escape velocities \citep[$2.6$ for OB and $1.6$ for classical Wolf-Rayet (cWR) and helium stars][]{Abbott1978,GrafenerVink2013}, $\Gamma\id{es}$ denotes the electron-scattering Eddington factor
\begin{equation}
\Gamma\id{es}=\frac{0.2(1+X) L}{4\pi c G M},
\end{equation}
$G$ is the gravitational constant, $c$ the speed of light and 
\begin{equation}
f\id{rot}=\sqrt{1-\left(\frac{\upsilon\id{rot}}{\upsilon\id{crit}}\right)^2}
\label{eq:Stellar_rotation_damping}
\end{equation}
\citep{Puls2008}.

Fig.~\ref{fig:Methods_Single_Initialmass_Energy} shows the kinetic energy $E_\mathrm{k}$
of the stellar wind integrated over the lifetime of the stellar models as a function of the initial mass $M_\mathrm{ini}$ for all the simulated rotation velocities and for $Z=0.004$. As expected, $E_\mathrm{k}$
increases with $M_\mathrm{ini}$. Stellar models with $M_\mathrm{ini}\approx\;$\M{100} eject around \SI{e51}{erg}, nearly three orders of magnitude more than stars with $M_\mathrm{ini}\approx\;$\M{8}. 
Stellar rotation
leads to higher values of $E_\mathrm{k}$. For instance, all the models
with $\upsilon_\mathrm{ini}=\;$\SI{600}{\kilo\meter\per\second} and $M_\mathrm{ini}>\;$\M{40} 
eject more than \SI{e51}{erg} in stellar winds.

In Fig.~\ref{fig:Methods_Single_Initialmass_Energy}, we use colours to highlight the relative contributions of different evolutionary stages\footnote{\label{footdef}We differentiate between the evolutionary phases based on the position of the models in the Hertzsprung–Russell diagram (HRD) and the optical depth of their winds. 
We identify stellar models in the upper right corner of the HRD as BSG/LBV stars and cool (i.e. with effective temperature $T_{\rm eff} \leq 10^4\,$K), lower luminosity stellar models as RSG. For the other types, we separate according to their position relative to the zero-age main sequence (ZAMS). Stellar models hotter than the ZAMS are either classified as helium stars or cWR stars, where cWR stars have optically thick winds (i.e. $\tau > 2/3$). Stellar models cooler than the ZAMS are either OB or WNL stars, where WNL stars have optically thick winds.
} to $E_\mathrm{k}$, namely
OB dwarfs (OB), red supergiants (RSG), blue supergiant/luminous blue variables (BSG/LBV), WNL stars (WNL), helium stars (He-star) and cWR stars. The corresponding discussion of the mass yields is provided in Appendix~\ref{sec:Appendix_massyield}.
For non-rotating models with $M_\mathrm{ini}\lesssim\;$\M{40},
a large fraction of the wind energy is
ejected during the main-sequence, when the stellar model would appear spectroscopically as a OB dwarf, with terminal wind velocities of the order of \SI{1000}{\kilo\meter\per\second} and mass-loss rates in the range \num{e-8}--\num{e-5}~\si{\Msun\per\year} \citep{vink2021r}. Stars in this mass range evolve past their main-sequence into red giants and supergiants.
However, despite the relatively large mass-loss rates, the contribution from the post-main-sequence evolution 
accounts only up to around $\sim10$ per cent of the total, due to the much slower winds of this phase \citep[of about \SI{10}{\kilo\meter\per\second},][]{deJager1988,Smith2014}. 
On the other hand,
stellar models with $M_\mathrm{ini}>\;$\M{40} lose their H-rich envelope and evolve into cWR stars
displaying dense and optically thick winds, with mass-loss rates of $\approx\;$\num{e-5}--\num{e-4}~\si{\Msun\per\year} and terminal velocities of $\approx\;$\SI{2000}{\kilo\meter\per\second} \citep{Nugis2000,Crowther2007,Smith2014}. Although
the WR phase lasts for $\approx 10$ per cent of the stellar lifetime for models with $M_\mathrm{ini}\approx\M{60}$, 
it accounts for about half of $E_\mathrm{k}$.
At even higher masses ($M_\mathrm{ini}\approx \M{100}$), 
the WR phenomenon already occurs during the main-sequence evolution, i.e. the stellar models correspond to WNL (or WNH) stars \citep{Crowther2007,Smith2014}. The relative importance of the WR stage (starting $\approx \SI{e6}{\year}$ after zero age) grows for $M_\mathrm{ini}\gtrsim\M{40}$, while the blue supergiants (including a potential LBV phase) only contribute a few per cent of the total mechanical yield.

Rotation makes stellar cores larger while reducing the extent of the H-rich envelopes thus producing more massive post-main-sequence helium/WR stars. 
In consequence, both the WR contribution and 
$E_\mathrm{k}$ increase at fixed initial mass (especially for  $M_\mathrm{ini}\approx\;$\num{40}--\num{50}~\si{\Msun}). Mass-loss rates are also enhanced although with lower terminal velocities.

\subsection{Wind feedback from binary stars}
\label{sec:wind_feedback_binary}

\begin{figure*}
	\centering
	\includegraphics[width=1\textwidth]{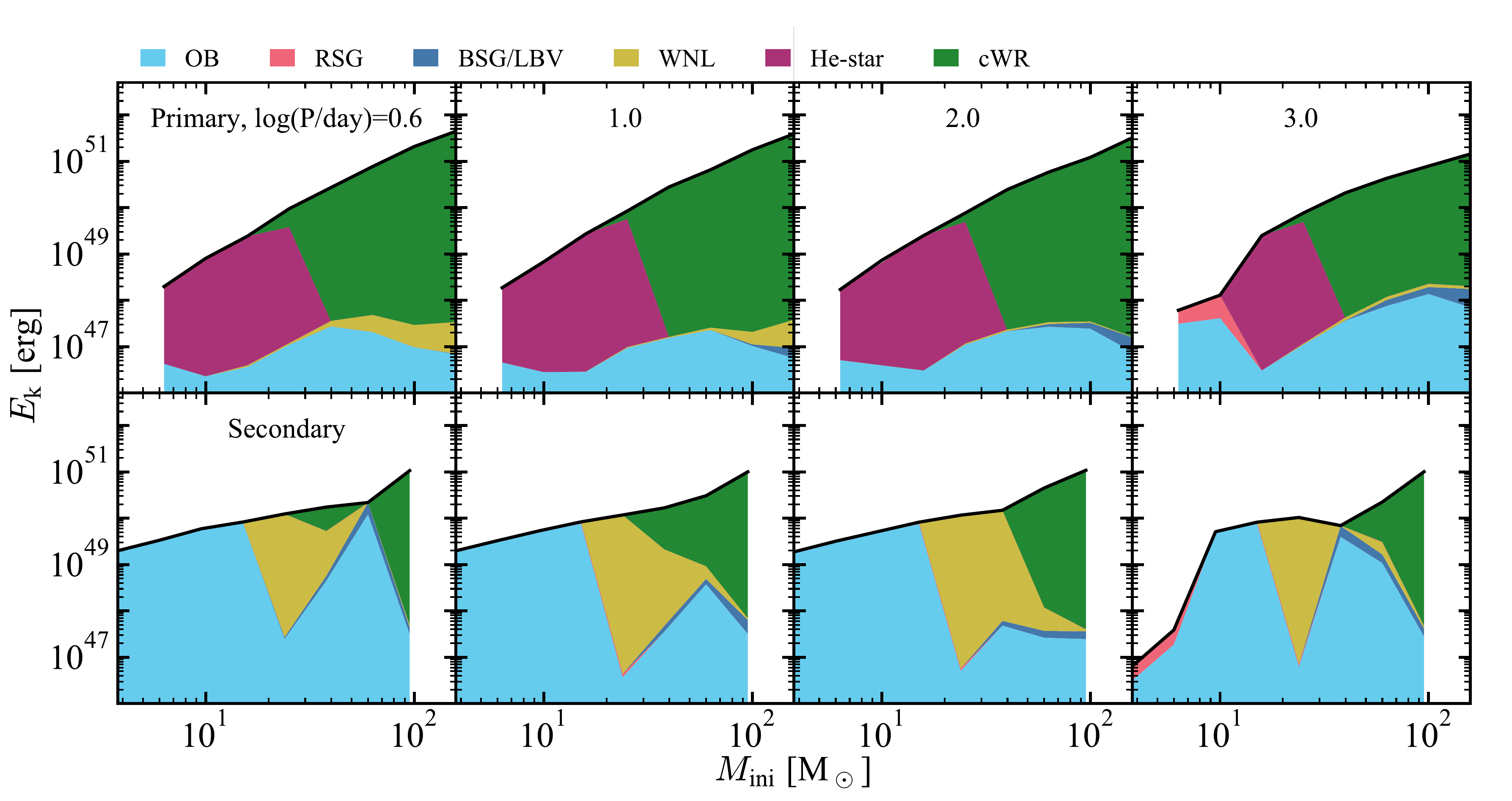}
	\caption{As in Fig.~\ref{fig:Methods_Single_Initialmass_Energy} but for binary systems with a fixed mass ratio of 0.6 and different periods (indicated by the labels). 
	}
	\label{fig:Methods_Binary_Initialmass_Energy}
\end{figure*}

It is well known that most stars form in binary systems \citep[e.g.][]{Sana2012} and this dramatically affects their evolutionary path  \citep{Podsiadlowski1992,VanBever2000,deMink2013,Mandel2016,Marchant2016,Langer2020}.
However, many aspects ranging from the way stars in binaries interact \citep[including common-envelope phases][]{Paczynski1976proceeding,Ivanova2013} to the intrinsic fraction of binary systems at different metallicities are still poorly understood. None the less, we attempt to investigate the impact of binaries to the expected mechanical and chemical feedback by stellar winds of stellar populations.

We compute binary stellar models with different metallicities (\numlist{0.0001;0.0004;0.0008;0.004;0.008;0.02}). For simplicity, we assume a fixed binary mass ratio of \num{0.6} and cover the same initial-mass range for the primary star as we did for single stars, although this time we use steps of $\log(M\id{ini}/\si{\Msun})=0.2$.
We consider four different
values for the orbital period, namely $P= 10^{0.6}, 10^{1}, 10^{2}$ and $10^3$ days. Finally, we take into account the fact that, in binary systems, some of the kinetic energy of the stellar winds is dissipated due to the 
interaction between the winds generated by the primary and secondary stars \citep{Usov1991,Stevens1992}. We base our estimate on the momentum balance between the stellar winds by adopting the reduction factor
\begin{align}
f\id{ww}=
\begin{cases}
1-\frac{\pi^2}{16}\sqrt{\frac{\dot{M}\id{s} \upsilon\id{inf,s}}{\dot{M}\id{p} \upsilon\id{inf,p}}}\quad &\text{for } \dot{M}\id{p} \upsilon\id{inf,p} > \dot{M}\id{s} \upsilon\id{inf,s},\\
1-\frac{\pi^2}{16}\sqrt{\frac{\dot{M}\id{p} \upsilon\id{inf,p}}{\dot{M}\id{s} \upsilon\id{inf,s}}}\quad &\text{otherwise },
\end{cases}
\label{eq:Stellar_energy_binary_damp}
\end{align}
\citep{DeBecker2013} where 
the indices p and s refer to the primary and secondary star, respectively. 
The maximum attenuation ($f\id{ww}=0.38$) is obtained 
when the two stars produce winds with equal momenta.

Once the primary star reaches core-helium exhaustion, we follow
the remaining evolution of the secondary as if it was a single star.
In small corners of the investigated parameter space, our evolutionary models suggest that a merger of the companions should take place, and the calculation is inevitably stopped as, presently, no stellar-evolution code is able to model such merger events \citep[see however][]{Glebbeek2013,Schneider2016}.
In these cases, we interpolate through our grid of models. 

In Fig.~\ref{fig:Methods_Binary_Initialmass_Energy}, we show $E_\mathrm{k}$ as a function of $M_\mathrm{ini}$
for the primary and secondary models.
Barring the most massive stars, both the primaries and the secondaries return more kinetic energy via their stellar winds compared to the single non-rotating case.
The largest contribution to $E_\mathrm{k}$ 
comes from 
helium and WR stars. In fact, all the binary models experience mass-transfer and 
the primaries 
loose their outer H-rich envelope. They thus achieve effective temperatures of about 
\num{5e4}--\num{e5}~\si{K}, becoming either stripped He-stars with optically thin winds or cWR stars, with much larger terminal wind velocities than late-type stellar models. The mass-gainers secondary models also show larger $E_\mathrm{k}$ compared to single star models of the same mass, particularly for the lowest considered initial mass and the short periods (i.e. those undergoing mass-transfer during the main-sequence phase). The reason is twofold: mass accretion on to the secondary models leads to higher masses and luminosities compared to the single star case, and a significant fraction of the mass stripped from the primary component of the binary systems is not accreated and leaves as stellar wind of the secondary model during its OB phase (given that mass-transfer takes place almost always during the main-sequence phase of the secondary stellar models), contributing to the kinetic stellar feedback (see also 
Appendix \ref{sec:Appendix_massyield}).

\subsection{Feedback from a stellar population}
\label{sec:fdbk_stel-pop}

In order to model the energy injection into the ISM by a simple coeval population of single stars, we use the parametrization for the initial mass function (IMF) introduced by \citet{Kroupa2001}.
We consider stars within the mass range
$10^{-1}\leq M \leq 10^{2.2}$ M$_\odot$
and account for the SN explosion and stellar winds of the models with $M>10^{0.9}$ M$_\odot$ but we assume that stars with $M>40$ M$_\odot$ collapse directly to a black hole at the end of their lifetime without releasing energy into the ISM.

For consistency with previous work, we associate an energy of \SI{e51}{erg} to every SN explosion \citep[e.g. ][]{Scannapieco2012, Agertz2013, Crain2015}, 
although this simplification
does not account for the variety of SNe seen in nature. 
We make sure that this energy is released into the ISM when a massive star reaches the end of its lifetime (i.e. the injection time is different for each stellar model).
We estimate the ejected mass by computing the
difference between the final mass of the stellar model and the average remnant mass derived in
\citet{Sukhbold2016}. 

We consider three cases. In the first one, we only use single non-rotating stellar models while, in the second one, we combine single rotating models according 
to an empirically derived distribution of initial rotation velocities which depends on metallicity (see Appendix~\ref{sec:Appendix_rotation}).
Finally, our third option also accounts for binary stars. In this case,
we assume that 70 per cent of the stellar mass is in binary systems and that the remaining 30 per cent is contributed by single stars.
For the binaries, we apply the Kroupa IMF to the mean mass in each system 
and assume a flat orbital-period distribution in log-space (between $\log (P/\mathrm{day})=0.3$ and 3.3) across all considered metallicities. 
This is consistent with observations
in both the Galaxy and the LMC \citep[][]{Kobulnicky2014,Almeida2017}.

\begin{figure}
	\centering
	\includegraphics[width=1\columnwidth]{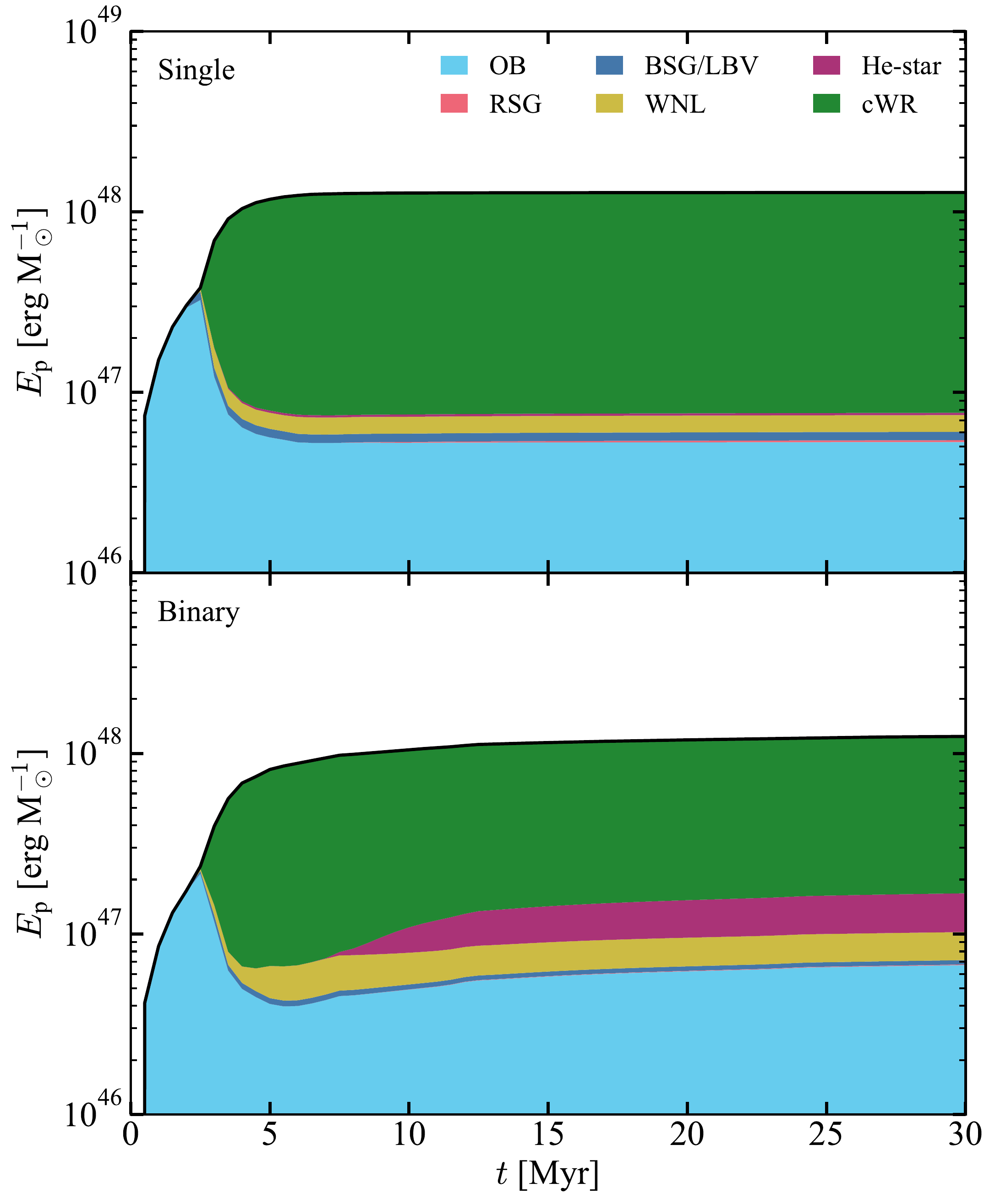}
	\caption{Cumulative kinetic energy ejected in the form of winds by a coeval simple stellar population with metallicity $Z=0.004$. Plotted is the energy per unit stellar mass as a function of the population age. The top and bottom panels refer to populations of single and binary stars, respectively.
	Colour shading is
	as in Fig.~\ref{fig:Methods_Single_Initialmass_Energy}.
	}
	\label{fig:Methods_Time_Energy}
\end{figure}

Fig.~\ref{fig:Methods_Time_Energy} shows the time dependence of the cumulative kinetic energy
ejected by one solar mass of coeval stellar evolutionary models (with $Z=0.004$) in the form of winds (see Appendix~\ref{sec:Appendix_massyield} for the corresponding discussion of the mass yields). The top panel refers to single rotating stellar models while binary systems are considered in the bottom one.
In the first \SI{2.5}{\Myr}, a few $\times\;$\SI{e47}{\erg\per\Msun} are
ejected by stellar models in the OB stage.
Subsequently, the most massive stellar models reach the WR phase, leading to a steep increase in the total energy ejected. Overall, the cWR stage contributes nearly 60 per cent of the total ejected energy (although only the most massive stars experience it). 
For single stars (binaries), 90 (65) per cent of the wind energy is ejected before the onset of the first SN explosion, which takes place when the stellar population has an age of \SI{5}{\Myr}. 
It is worth mentioning that the binary systems
show more prominent contributions from the post-main-sequence helium star models. For instance, stripped helium stars originating from binary interactions start playing a relevant role after \SI{10}{\Myr} and provide roughly 10 per cent of the total energy.

\begin{figure}
	\centering
	\includegraphics[width=1\columnwidth]{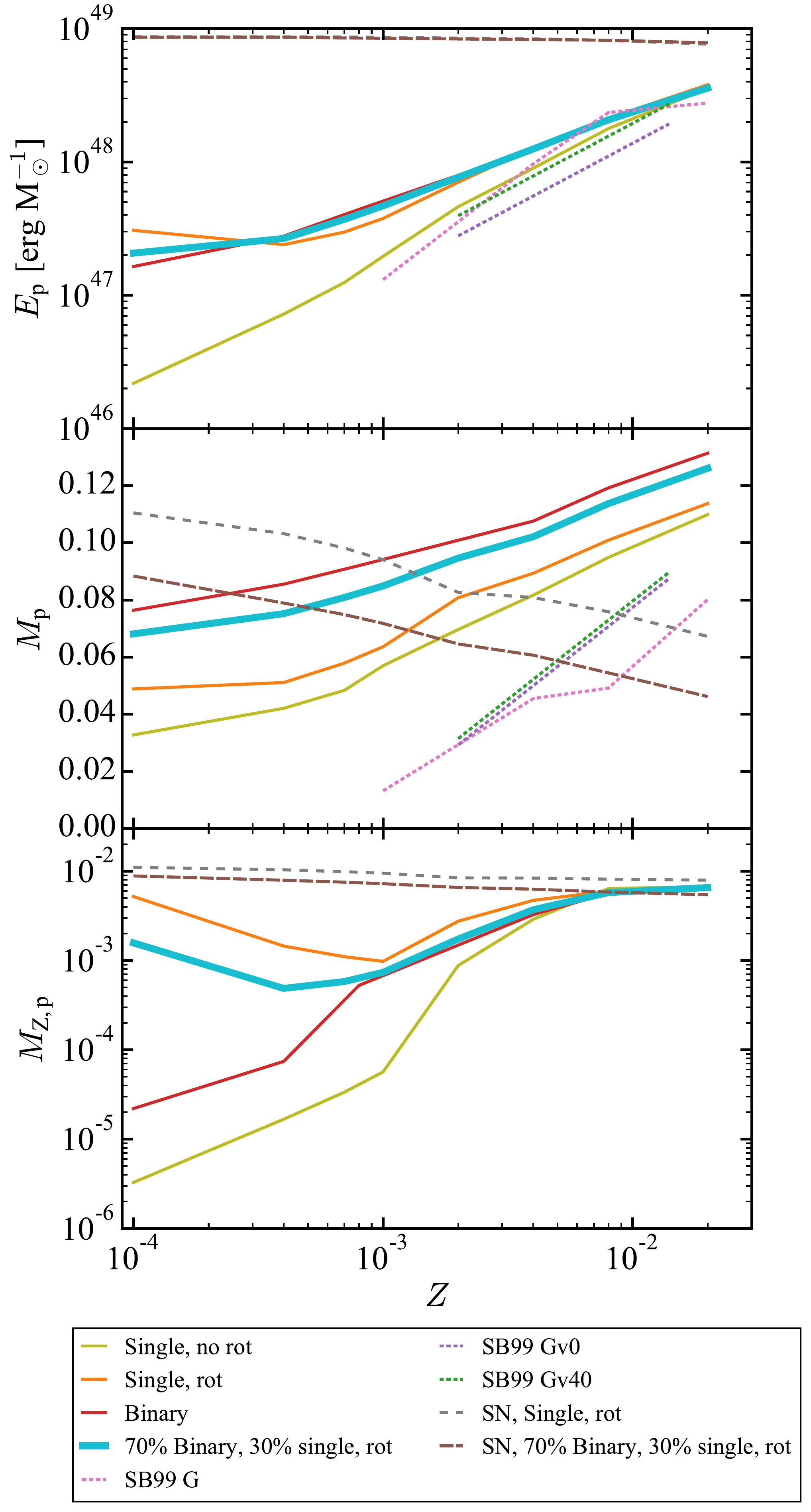}
	\caption{Mechanical and chemical yields of winds emitted by different stellar populations (see labels) within the first \SI{30}{\Myr} 
	as a function of absolute metallicity. 
	From top to bottom in the panels, we show
	the (time-integrated) ejected energy per unit stellar mass, the ejected mass fraction and the corresponding metal fraction.
	For comparison, the corresponding yields of SNe are shown with dashed lines (assuming the metal yield 
	is 10 per cent). Also shown are the yields of three different tracks from \textsc{Starburst99}: Geneva standard \citep[SB99 G,][]{Schaller1992}, v00 \citep[SB99 Gv0,][tracks with $\upsilon_\mathrm{ini}=0$]{Ekstrom2012} and v40 \citep[SB99 Gv40,][tracks with 
	$\upsilon_\mathrm{ini}=0.4\upsilon\id{crit}$]
	{Ekstrom2012}.
	}
	\label{fig:Methods_final}
\end{figure}

So far we have only discussed the stellar models with $Z=0.004$. In Fig.~\ref{fig:Methods_final},
we present the metallicity dependence of the kinetic energy ($E_\mathrm{p}$, top panel), mass ($M_\mathrm{p}$, middle panel), and metals ($M_{\mathrm{Z, p}}$, bottom panel) ejected from a coeval population per unit stellar mass.
Single non-rotating stars show the strongest metallicity dependence of $E_\mathrm{p}$ and $M_\mathrm{Z, p}$ while the inclusion
of rotating and binary stars leads to a remarkably shallower relation. The reason is twofold: (i) the probability density function of the initial rotational velocity peaks at higher velocities with decreasing $Z$
(thus driving larger rates of mass-loss by stellar winds) and (ii) binary interaction leads to the formation of stripped helium and WR stars also when envelope self-stripping by stellar winds is not sufficient, i.e. at low $Z$. 
Note that, at low metallicity, single non-rotating stars eject nearly one order of magnitude less
energy and three orders of magnitude less metals
than a more realistic population composed of a mix
of rotating and binary stars. This difference
could play an important role in modelling the early phases of galaxy formation.
Regarding the integrated mass-loss quantified by $M_\mathrm{p}$, we note that
all model populations show a similar metallicity
dependence although with systematically higher mass yields from the rotating and binary models, by approximately a factor two.

We compare our results for a coeval stellar population including a combination of binary systems and rotating single stars with the results from \textsc{Starburst99}, which are derived adopting the Geneva evolutionary tracks for single stars. 
We use the same IMF but the most massive model in the Geneva set
(\M{120}) is lower than in our set.
For the non-rotating single stars, the ejected energy per unit stellar mass we derive is consistent with that from \textsc{Starburst99}, with an approximate metallicity dependence of $E\id{p} \propto Z$. On the other hand, the mix of 
binary systems and rotating single-star models shows a shallower relation with $E\id{p} \propto Z^{0.65}$.
\subsection{Discussion and a caveat}
\label{sec:fdbk_discussion}

Stars inject kinetic energy and mass into the surrounding medium during their whole lifetime.
However, a fraction of this energy is likely dissipated in the nearby circumstellar regions \citep[e.g.,][]{GarciaSeguraMacLow1996,GarciaSeguraLanger1996} due to the variable wind velocity over the evolution of a single star \citep{Vink2001}.
For instance, the slow and dense RSG or LBV outflows swept by the faster wind during the following WR phase can lead to a significant fraction of the kinetic energy being lost at scales smaller than the minimum resolution of our simulations of galaxy formation ($\sim\SI{40}{pc}$, see Section~\ref{sec:numerical_methods}).
Moreover, in a stellar cluster, some energy will be also dissipated in the shocks forming 
between colliding stellar winds and between the winds and the clumpy ISM. As mentioned in the introduction,
different authors reach opposite conclusions regarding the entity of the dissipation \citep{Rey-Raposo2017, Lancaster2021, Rosen2021}. In the absence of a consensus, in this work, we inject the whole energy released by the winds and SNe into the ISM 
and solve the equations of fluid dynamics to determine 
gas flows on scales of tens of pc.
In this sense, our study quantifies the maximum effect that could be possibly driven by stellar winds.

It is also important to stress that, 
despite the large observational and theoretical efforts, 
mass-loss prescriptions in stellar models are still uncertain\footnote{This is mostly due to the presence of inhomogeneities, called clumping, that affect the line diagnostics \citep{Moffat1988,Puls2008,Sundqvist2013}.} by about a factor of three at Galactic metallicity \citep[][and references therein]{Smith2014,vink2021r}. Uncertainties are even more severe at lower metallicities, where the lack of observational constrains and the low abundance of the metals 
that are responsible of
the radiatively driven stellar wind do not allow for fully reliable estimates. 
This is particularly the case for the cWR stars which, as shown in Fig.~\ref{fig:Methods_Single_Initialmass_Energy} produce a large amount of mechanical feedback. 
These uncertainties present a major challenge in determining the energy and momentum budget that a stellar population injects into the ISM in the form of winds.

\section{Simulating stellar feedback}
\label{sec:sim_fdbk_over}
\subsection{The state of the art}
Different methods have been used to include SN feedback in numerical hydrodynamic simulations of galaxy formation. In cosmological runs that cover large volumes and achieve a spatial resolution of $\simeq 0.2$--$1$ kpc (insufficient to reveal individual SNRs and the complex multi-phase structure of the ISM), SN feedback was originally implemented as a single injection of thermal energy and mass from a simple stellar population.
It turns out, however, that the energy is deposited in too large a volume or mass. Consequently, the bulk of the injected energy is immediately radiated away (without having much mechanical impact on the ISM) due to the high densities of the star-forming regions \citep[e.g.][]{Katz1992}.
In order to prevent this `overcooling problem' some ad hoc shortcuts have been adopted.
One possibility is to artificially prevent the heated gas from cooling for a time comparable with the local dynamical time-scale ($\sim30$ Myr) so that to convert part of the deposited energy into actual gas kinetic energy and mimic the production of hot bubbles (possibly driving outflows) generated by the combination of blast waves from multiple SN explosions \citep{Gerritsen1997PhDT, GerritsenIcke1997, Thacker2000,  SommerLarsen2003, Keres2005, Stinson2006, Governato2007}. The rationale behind this method is that the volume-filling cavities of low-density gas in the multi-phase ISM are not resolved and thus radiative losses are overestimated in the simulations.
Alternatively, one can implement a `kinetic feedback' scheme in which
the gas elements are imparted
some outwardly directed momentum (and
are decoupled from hydrodynamic forces until they leave the galaxy) to simulate the launching of a galactic wind 
\citep[e.g.][]{NavarroWhite1993,MihosHernquist1994, SpringelHernquist2003, Oppenheimer2010, Vogelsberger2013, Dave2017, Valentini2017, Pillepich2018}.
This can be implemented in many different ways and generally requires the introduction of free parameters. \citet[][see also \citealt{Mori1997}, \citealt{Gnedin1998}]{Dubois2008}, for instance, 
impose around each SN event
a spherical Sedov blast-wave profile for density, momentum and total energy, with a radius equal to a fixed physical scale resolved with a few computational elements. This radius should approximately match the size of super-bubbles blown by multiple clustered SN explosions (hundreds of pc) and is obviously much larger than individual SNRs. Note that this radius also sets the injection scale for turbulence in the simulated ISM.
Kinetic feedback schemes, however, do not properly account for the thermal state of the ISM as they neglect the hot phase produced by SNe.
Approximate sub-grid models are thus used to determine the `effective pressure' of the multi-phase ISM as a function of the coarse-grained
density of the gas in the simulations 
\citep{Yepes1997, HultmanPharasyn1999, SpringelHernquist2003, SchayeDallaVecchia2008, Murante2010}. Basically, the use of an effective
equation of state prevents the dense
gas from cooling down to arbitrarily small temperatures and artificially fragmenting.

On the opposite extreme, non-cosmological simulations of either an idealized ISM or individual molecular clouds and parts of disc galaxies have investigated the impact of individual SNe on their surroundings
with sub-pc spatial resolution
\citep[e.g.][]{Thornton1998, Creasey2013, Gatto2015, IffrigHennebelle2015, KimOstriker2015, WalchNaab2015, Walch2015, Girichidis2016, Simpson2016, KimOstriker2017, Hirai2021, Smith2021}.
The emerging picture is that
most of the energy injected by a SN
into the ISM 
is rapidly thermalised and radiated away. However, a fraction ranging
between a few to ten per cent 
is later found as kinetic energy of the ambient medium, with some dependence on the local density and the time at which the retained energy is estimated.

In the last decade or so, it has become computationally feasible to simulate
individual galaxies with a spatial resolution of a few tens of pc which partially reveals the turbulent and multi-phase structure of the ISM. 
At this resolution, the largest sites of star formation (where giant molecular clouds come into existence) can be localised in the simulations although their internal structure cannot be probed.
\citet{CeverinoKlypin2009} showed that, with a spatial resolution of $\sim50$ pc, it is possible to drive galactic winds without artificially delaying gas cooling after injecting SN energy.
This, however, requires that some SNe explode outside of the dense regions in which they formed due to OB-runaway stars.
Several studies devised an optimal strategy for simulating SNR and deal with the overcooling problem \citep[][but see also \citealt{Hopkins2014} and \citealt{KimmCen2014}]{KimOstriker2015, Martizzi2015}.
If the radius of the shell forming at the end of the Sedov-Taylor stage (which is sometimes called the cooling radius) is resolved with a sufficient number of elements ($\sim10$), then one should inject $10^{51}$ erg of thermal energy and let the hydrodynamic solver track the buildup of momentum. Otherwise, if the ambient density is too high for the achieved resolution (and thus the shell radius too small), one should directly inject the full momentum generated during the Sedov-Taylor phase. The latter approximation fails to catch the impact of the hot gas (and, possibly, underestimates galactic winds) but anyway drives turbulence in the warm and cold phases of the ISM and allows for self-regulated star formation.

\subsection{Our implementation}
We adopt two different numerical schemes to simulate SN feedback.
In the first one (dubbed T as a short for `thermal feedback'),
the mass, metals and energy ejected by SNe are deposited
at the location of stellar particles with age $t\id{SN}$. 
The SN energy is used to alter the thermal budget of the surrounding ISM
and 
gas cooling is switched off as in the standard \textsc{Ramses} implementation
with a characteristic time-scale of 20 Myr.

In the second scheme (dubbed M as a short for `mixed thermal-kinetic feedback'),
the ejecta are distributed within a sphere of radius $r\id{Sed} = 150$ pc
and the SN energy is partitioned between the thermal energy of the gas (corresponding to 70 per cent of the total) and a kinetic term accounting for the bulk motion of the ISM
within $r\id{Sed}$ \citep{Dubois2008}. 
Mass, momentum and kinetic energy are distributed within $r\id{Sed}$ 
as in a spherically-symmetric Sedov blast wave.
We assume a unitary mass-loading factor, i.e. that the gas mass entrained
by the SN explosion is equal to that of the stellar particle.
Delayed cooling is implemented as in the T model.

We use the 
blast-wave model also to describe the combined effect of
stellar winds and SNe.
In this case, however, mass, energy and momentum are continuously injected
into the ISM by young stellar particles.
The corresponding rates are obtained by interpolating the tables we derived in Section~\ref{sec:wind_feedback}.

\section{Numerical methods}
\label{sec:numerical_methods}
As a first sample application of our stellar models to the field
of galaxy formation and evolution, 
we investigate the impact of stellar winds on the structure of a  
high-redshift galaxy.
To this end,
we use the adaptive-mesh-refinement (AMR) code \textsc{Ramses} \citep{Teyssier2002} to
perform several cosmological zoom-in simulations 
of a
sub-$L_*$ galaxy at $z=3$. Each simulation starts from the same initial conditions (IC)
but adopts different feedback models.

\subsection{Initial conditions and refinement strategy}
\label{sec:ics}
We consider
a flat Lambda cold dark matter cosmological model 
\citep{Planck2020} 
and simulate the formation of a galaxy
within a cubic periodic box of comoving side $L_\mathrm{box}= 6 ~\hmpc$.
Initial conditions (IC) are generated at $z=99$ with the \textsc{Music} code \citep{Music2011}.
We set up zoom-in simulations following a multistep procedure. 
We first generate IC 
with a uniform resolution 
using $2^{l\id{ini}}$ elements along each spatial dimension
(with $l\id{ini}=7$)
and run a DM-only simulation until $z=3$.
After identifying haloes
with the \textsc{Amiga Halo Finder} code \citep[\textsc{ahf},][]{Gill2004,Knollmann2009},
we pick a halo with a virial mass of $M\id{vir} \sim10^{11} \Ms$ and a relatively quiescent late mass-accretion history (the virial radius $R\id{vir}$
encloses a mean density of $200\,\rho\id{crit}$, with $\rho\id{crit}$ being the critical density for closure of the Universe).
At this point, we use
the zoom-in technique to re-simulate the selected halo
at higher spatial resolution.
In brief,
all the simulation particles found within 
$3R\id{vir}$ from the halo centre at $z=3$ are traced back to the IC and the corresponding volume is 
resampled at a higher resolution. 
The final configuration includes several nested levels with different spatial resolutions.

We investigate discreteness effects and numerical convergence (see Sections~\ref{sec:results_globalproperties} and \ref{sec:comp_vs_obs.}) by generating ICs with two different
maximum levels of refinement ($l\id{ini}=9$ and 10)
corresponding to different mass resolutions for the DM component
($m_\mathrm{DM}=1.8 \times 10^5$ and $2.2 \times 10^4$ M$_\odot$, respectively).
In the remainder of this paper, we refer to the simulations
obtained in the two cases as `low-' and 'high-resolution'.

During run time, the AMR technique refines (and coarsen) the grid structure used to solve
the fluid equations for the gas and the Poisson equation for the gravitational potential based on pre-determined criteria.
We adopt a quasi-Lagrangian scheme which is uniquely based on mass density.
A cell is split if it contains more than eight DM particles
or a total baryonic mass (stars and gas) corresponding to the same density contrast.
In order to prevent runaway refinement early on, 
triggering a new level is only allowed after the simulation has reached a
specific cosmic time so that to match the refinement strategy
of the corresponding DM-only case (which we always run first). 
This strategy makes sure that the 
resolution of the grid in physical
units remains nearly constant (in a stepwise sense).
For all simulations, the AMR algorithm adds six levels of refinement
until $z=3$, corresponding to a nominal spatial resolution of $68$ and $34$ physical pc for the low- and high-resolution simulations, respectively.

In order to facilitate the interpretation of the results,
we save snapshots every 5 Myr for the simulations with $l\id{ini}=10$
(10 Myr for those with $l\id{ini}=9$).

\subsection{Gas physics and star formation}
\label{sec:sf}
For the gas,
we assume an equation of state with polytropic index $\gamma =5/3$. 
In order to avoid spurious fragmentation, we artificially add thermal pressure
where needed in order to resolve the Jeans length with four grid elements or more
\citep{Truelove1997,Teyssier+10}.
The gas is ionised and heated up by 
the default
time-dependent  
uniform cosmic UV background in \textsc{Ramses} 
which is exponentially suppressed where the physical number density of
gas particles, $n$, exceeds $0.01\, \mathrm{cm}^{-3}$ to approximate self-shielding
\citep[e.g.][]{Tajiri-Umemura1998}.
Our runs include gas cooling from H, He and metals. 

The conversion of `cold' and dense gas into stars is modelled with a Poisson process
in which the average follows the relation \citep{Schmidt1959}
\begin{equation}
\dot{\rho}_\star=\frac{\epsilon_\star}{t\id{ff}}\,\rho\id{gas}\,,
\ \ \ \ \text{if}\ T<\SI{2e4}{\kelvin}\ \ \text{and} \ \ n>n_\star\,,
\label{eq:Theory_sf}
\end{equation}
which relates the
star-formation-rate (SFR) density
$\dot{\rho}_\star$ and the gas density
$\rho\id{gas}$ in terms of 
the star-formation efficiency $\epsilon_\star= 0.01$
and the free-fall time of the gas
$r\id{ff}=\sqrt{3\pi /(32G\rho\id{gas})}$ 
\citep[e.g.][]{Agertz2011,Perret2014,Kretschmer2020}. 
We set the density threshold for star formation to $n_\star=10$
and $20\,\mathrm{cm}^{-3}$ in the low- and high-resolution
simulations, respectively.
Stellar particles have a mass of $2.5 \times 10^4$ M$_\odot$ in the high-resolution runs which is large enough to properly sample the IMF also for massive stars.
\begin{table}
\caption{Main properties of the simulation suite.}
\label{tab:Method_Simulations}
\centering
	\begin{tabular}{ccccccc} \hline
		Name & Feedback & $t\id{SN}$ & Winds& $l\id{ini}/l\id{max}$& $\Delta x$ & $n_\star$ \\ 
	 & & Myr & & &  \si{\pc}& \si{\pcm}\\  \hline
		S9T & Thermal& 10 & No &\num{9}/\num{15} & \num{68} & \num{10}  \\
		S9M & Mixed& 10 & No &\num{9}/\num{15} & \num{68} & \num{10}  \\
		E9S & Mixed& Variable & No &\num{9}/\num{15} & \num{68}& \num{10} \\ 
		E9O & Mixed& Variable& Non-rot &\num{9}/\num{15} & \num{68} & \num{10} \\ 
		E9R & Mixed& Variable & Rot&\num{9}/\num{15} & \num{68}& \num{10} \\ 
		E9B & Mixed& Variable & Rot+Bin& \num{9}/\num{15} & \num{68}& \num{10}  \vspace{0.2cm}\\
		S10M & Mixed& 10 & No &\num{10}\num{16}&\num{34}& \num{20} \\
		E10R & Mixed& Variable & Rot &\num{10}/\num{16} & \num{34}& \num{20} \\
		E10B & Mixed& Variable& Rot+Bin&\num{10}/\num{16} & \num{34}& \num{20} \\ \hline
	\end{tabular}
\end{table}

\subsection{Stellar feedback}
\label{sec:sfdk}

We consider two sets of simulations. Namely, those adopting a
standard SN feedback model (name starting with S) and those 
based on our stellar tracks (name starting with E, short for early feedback).
Each stellar particle in the simulations represents a
coeval stellar population.
In the standard feedback model,
we consider type II SNe originating from these populations.
A single occurrence takes place $t\id{SN}= 10$ Myr \citep[]{Agertz2011, Ocvirk2020} after the birth of the stellar particle. 
The total ejected mass and energy 
by SNe
from a stellar population with mass $M\id{pop}$ are
 $M\id{ej}=0.1\,M\id{pop}$ 
and $E\id{ej}=(M\id{ej}/\SI{10}{M_\odot})\, \SI{e51}{erg}$ 
\citep{Dubois2008,Agertz2011,Ocvirk2020}
which is equivalent to assuming that every SN event injects (on average) $10^{51}$ erg and \SI{10}{M_\odot} into the ISM. 
The corresponding metal yield is 10 per cent \citep[see equation 4 in][]{Perret2014}. 

In the E-simulations, instead, SN events do not all take place after 10 Myr after
the birth of a stellar particle. On the contrary, mass, metals and energy are injected into the ISM at all times reflecting the lifetime of the stellar models
presented in Section~\ref{sec:wind_feedback} and the IMF by \citet{Kroupa2001}.
In this case, the masses ejected by SNe 
are
obtained from the stellar tracks (see Section~\ref{sec:fdbk_stel-pop})
while we still assume that each SN event releases $10^{51}$ erg and
that the metal yield is 10 per cent.
In addition, the contribution generated by stellar winds is also taken into account.

A list of all simulations and their distinguishing features are presented in Table~\ref{tab:Method_Simulations}. The naming convention is as follows: as previously mentioned, the first letter (S or E) indicates whether a standard
SN-only model or the more-sophisticated early-feedback scheme is used.
This is followed by a number which gives the maximum level of refinement in the IC (9 or 10). Finally, the last letter gives further details about the adopted feedback model.
For the S simulations, T and M stand for  
thermal feedback and mixed thermal-kinetic feedback, respectively.
On the other hand, for the E simulations (which all use the blast-wave
model) there are four possibilities. The letter S indicates that only
SN feedback is considered while the letters O, R and B refer to 
SNe and winds from non-rotating single stars, rotating single stars, and a mix of
rotating single stars and binaries, respectively.



\section{Results}
\label{sec:results}
\begin{figure}
	\centering    
	\includegraphics[width=1\columnwidth]{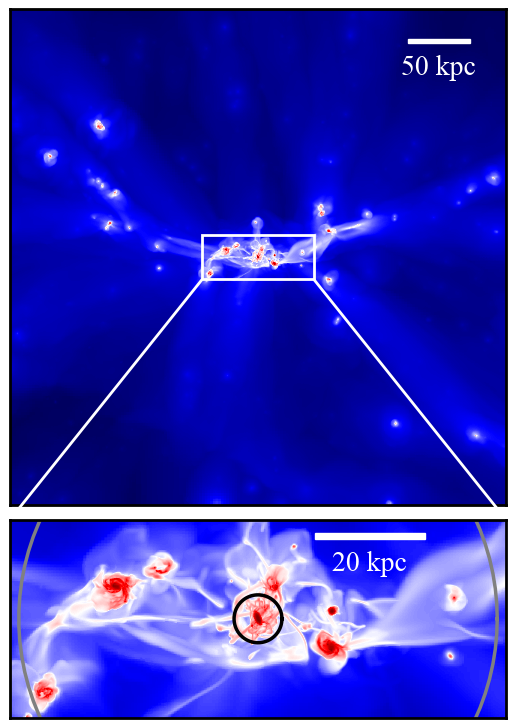}
	\includegraphics[width=1\columnwidth]{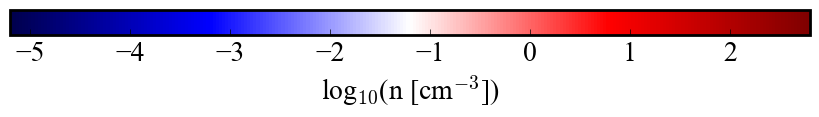}
	\caption
	{Projected map of the gas distribution in the E10B run at $z=3$.
	Each pixel is
	colour coded according to the maximum density along the line of sight
	within a cube of side length 400 kpc.
	In the bottom zoomed-in panel, the halo and the galaxy radii are indicated with a grey and a black circle,
	respectively. 
	}
	\label{fig:Results_Images_Evironment}
\end{figure}

We now present the results of the simulations. In what follows, we assume a solar metallicity of 2 per cent by mass (i.e. $\si{\Zsun}=0.02$) and all distances are given in physical units unless explicitly stated otherwise.

The virial mass and radius of the re-simulated DM halo at $z=3$ are nearly the same in all simulations, namely $M\id{vir}=1.8 \times 10^{11}$ M$_\odot$ and $R\id{vir}=43$ kpc. 
We conventionally define the central galaxy as the collection of gas and stars
enclosed within a radius of $R\id{gal}=0.1\,R\id{rvir}$ 
and located at the centre of the DM halo
\citep[e.g.][]{Scannapieco2012}. 
In order to give a visual impression of the environment
surrounding the galaxy, in Fig.~\ref{fig:Results_Images_Evironment}, 
we show a projected map of the gas density for the run E10B at $z=3$. 
The top panel has a side length of 400 kpc and illustrates 
the area of the intergalactic medium in which the DM halo resides
while the bottom panel zooms in the central region
(here, $R\id{vir}$ and $R\id{gal}$ are highlighted with
a grey and a black circle, respectively). 
The intricate habitat of the galaxy within $R\id{vir}$ is characterised by a nearly planar gas distribution
veined with dense filaments feeding the central object and its multiple companions (which, in all probability, will eventually merge with the main galaxy).

\begin{table}
	\caption{Properties of the simulated galaxies at $z=3$. 
	}
	\label{tab:gal_properties}
	\centering
	\begin{tabular}{cccccc} \hline
		Name & $M_\star$ &  $M\id{gas}$ & SFR & $Z\id{star}$ & $Z\id{gas}$  \\ 
		& \num{e9} \si{\Msun} & \num{e10} \si{\Msun} & \si{\Msun\per\year} &$\si{\Zsun}$  & $\si{\Zsun}$  \\
		\hline
		S9T  & \num{5.84} & \num{0.93} & \num{9.34} & \num{0.169} & \num{0.248}\\
		S9M  & \num{2.40} & \num{1.37} & \num{4.02} & \num{0.058} & \num{0.092}\\
		E9S  & \num{1.90} & \num{1.41} & \num{3.17} & \num{0.047} & \num{0.065} \\
		E9O  & \num{0.96} & \num{1.47} & \num{1.38} & \num{0.030} & \num{0.043}\\
		E9R  & \num{0.97} & \num{1.49} & \num{1.32} & \num{0.033} & \num{0.046}\\
		E9B  & \num{0.97} & \num{1.47} & \num{1.31} & \num{0.022} & \num{0.031} \vspace{0.2cm}\\	
		S10M & \num{3.80} & \num{1.25} & \num{5.11} & \num{0.087} & \num{0.142}\\
		E10R & \num{1.09} & \num{1.59} & \num{1.48} & \num{0.034} & \num{0.049}\\
		E10B & \num{1.16} & \num{1.55} & \num{1.49} & \num{0.024} & \num{0.035}\\ \hline 
	\end{tabular}
\end{table}

\begin{figure*}
	\includegraphics[height=0.15\textheight]{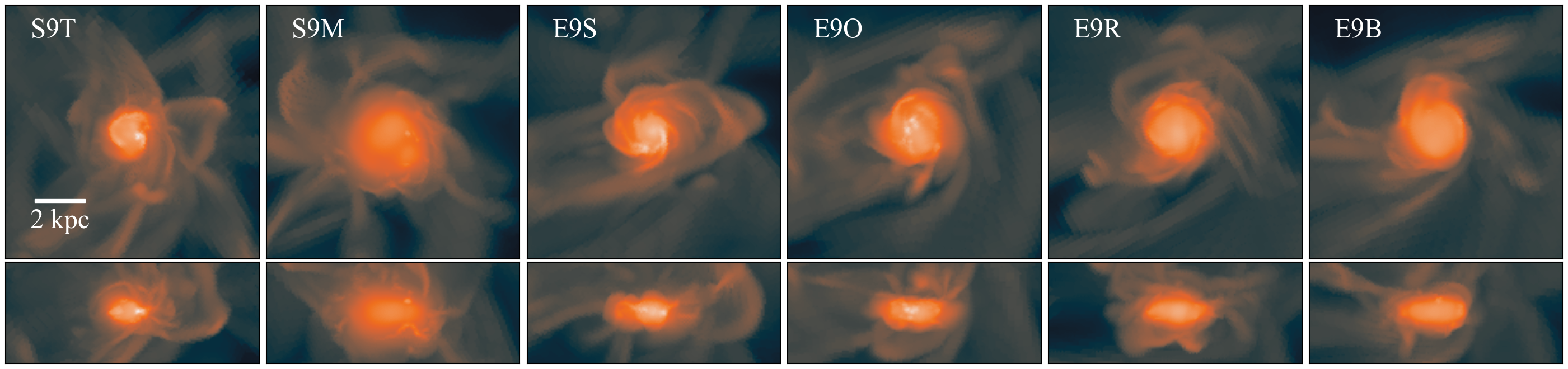}
	\includegraphics[height=0.15\textheight]{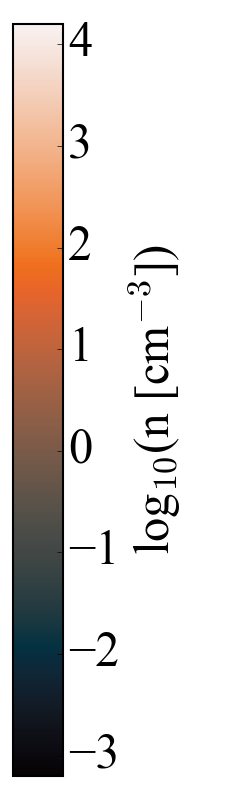}
    \includegraphics[height=0.15\textheight]{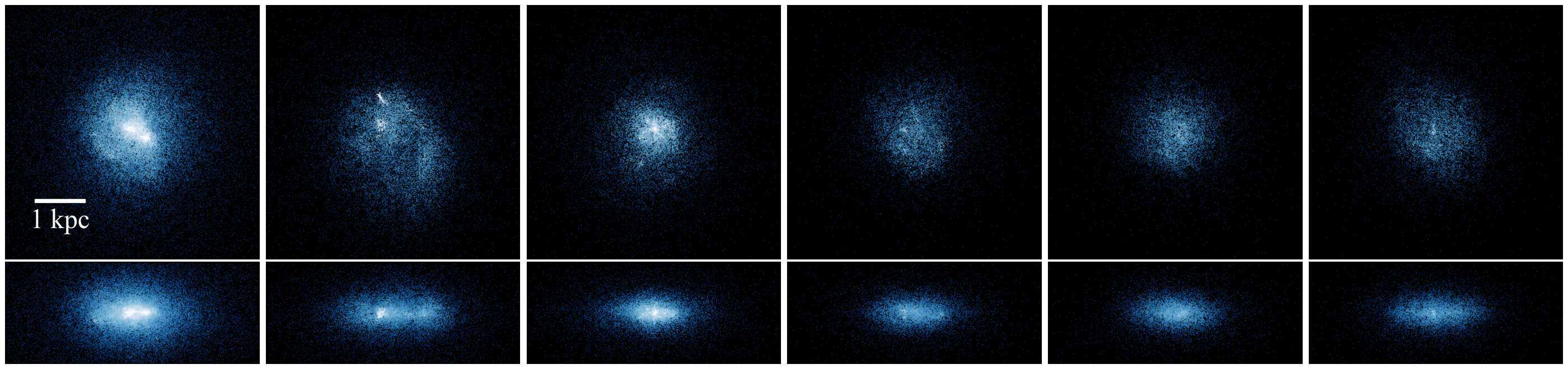}
   \hspace{-0.12cm} 
   \includegraphics[height=0.15\textheight]{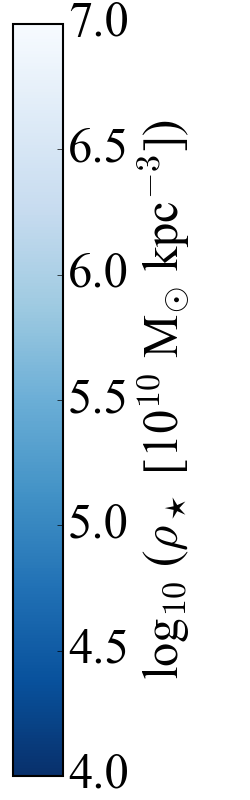} 
	\caption[Gaseous (top panels) and stellar (bottom panels) density components images for the low-resolution runs at $z=3$.]
	{Face-on and edge-on false-colour images of the gas (top row) and stellar (bottom row) distributions in the low-resolutions simulations at $z=3$.
	Shown are the maximum gas density along the line of sight and the star volume density
	at the location of the stellar particles.
	Note the different length scales used for gas and stars as indicated by the white yardsticks.
	}
	\label{fig:Results_Images_low}
\end{figure*}

\subsection{Global properties and morphology}
\label{sec:results_globalproperties}
The main properties of the simulated galaxies at $z=3$
are listed in Table~\ref{tab:gal_properties}.
These include the stellar mass ($M_*$), the gas mass ($M\id{gas}$), the average
SFR over the past \SI{200}{\Myr} as well as the mass-weighted
mean metallicity of the stars ($Z\id{star}$) and of the gas ($Z\id{gas}$),
both in solar units.

As a first step, we compare the different SN-only runs.
Looking at the low-resolutions simulations reveals that
the thermal feedback scheme (S9T) is rather inefficient as it produces nearly three times larger $M_*$,
$Z\id{star}$ and $Z\id{gas}$
than those obtained considering the two mixed schemes based on the blast-wave model (S9M and E9S).
On the other hand, considering that massive stars have different lifetimes
(E9S) causes a 20 per cent reduction in $M_*$ and $Z\id{star}$ with respect to S9M.

Accounting for stellar winds substantially reduces $M_*$, the SFR
and the metallicities (all by at least a factor ranging between 2 and 3.5) while it does not leave an imprint in $M\id{gas}$.
In fact, it is to be expected that
winds make stellar feedback more efficient (i) by injecting non-negligible amounts of energy and momentum into the ISM and (ii) by lowering the gas density around stellar particles before SN explosions take place.
However, the different stellar-wind models (O, R, B) appear to generate galaxies with very similar global properties although accounting for rotation
and binaries substantially increases the instantaneous energy injection rate.
This unexpected behaviour arises from the fact that the winds of
single non-rotating stars already provide enough energy early on to raise the gas temperature in the simulations above the threshold at which star formation is inhibited
 \citep{Stinson2013} -- see equation~(\ref{eq:Theory_sf}). 
Due to this threshold effect, the SFR
of the galaxies does not change much when additional energy is injected 
by rotating and binary stars.
Basically, when winds are accounted for, the formation of stellar particles in the spatial and temporal
proximity of recently born ones is reduced considerably with respect
to the simulations that only consider SN feedback.

Another thing worth mentioning is that the O, R and B wind models generate galaxies
with similar values of $Z\id{star}$ and $Z\id{gas}$
although the metal yields presented in Fig.~\ref{fig:Methods_final} can differ by orders of magnitude. This stems from the fact that the metal injection is dominated by SNe. 
Anyway, $Z\id{star}$ and $Z\id{gas}$ are nearly
30 per cent lower in the E9B and E10B simulations compared to the corresponding O and R
runs.
On the one hand,
the metallicities of our galaxies at $z=3$ are in very good agreement with those
found in other simulations for objects with similar stellar masses and using an early-feedback scheme \citep[e.g.][]{Ma2016}. 
On the other hand, they appear to be low compared to estimates based on observational data
\citep[e.g.][]{Sommariva2012,Arabsalmani2018} which, however, are known to suffer from
uncertain calibrations.

\begin{figure}
	\centering    
	\includegraphics[height=0.15\textheight]{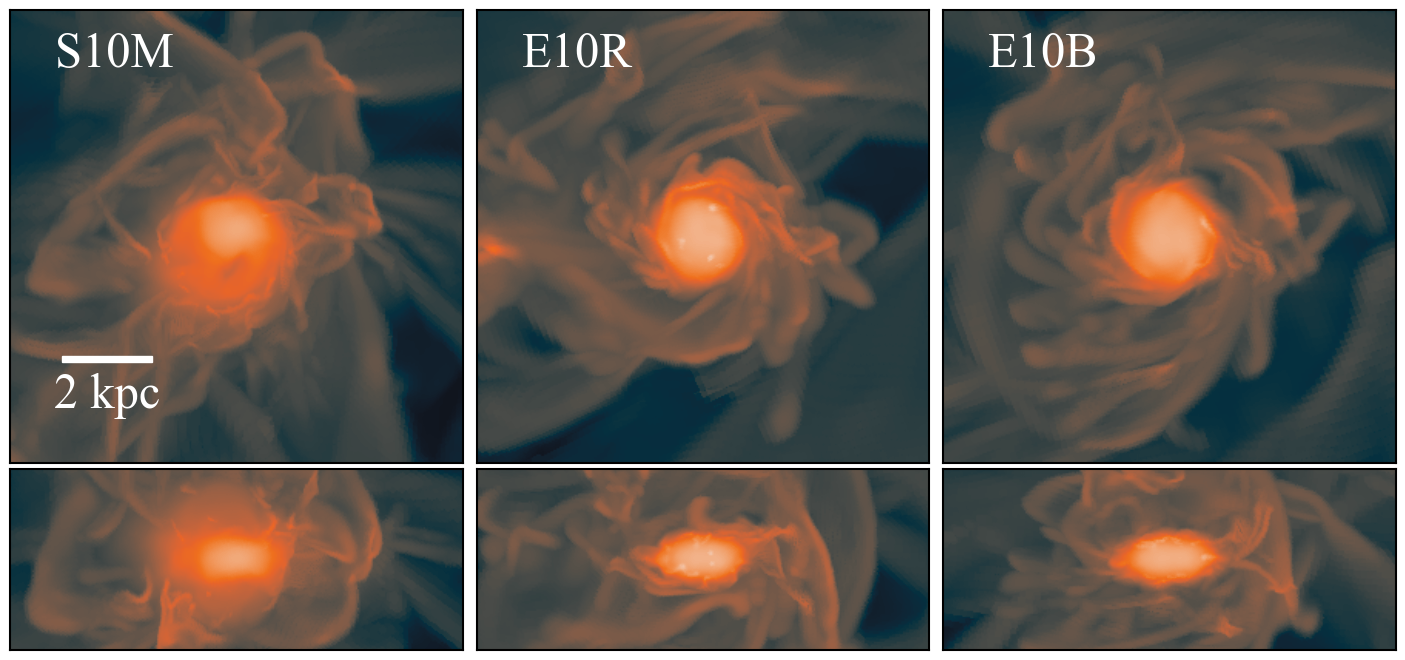}
	\includegraphics[width=1\columnwidth]{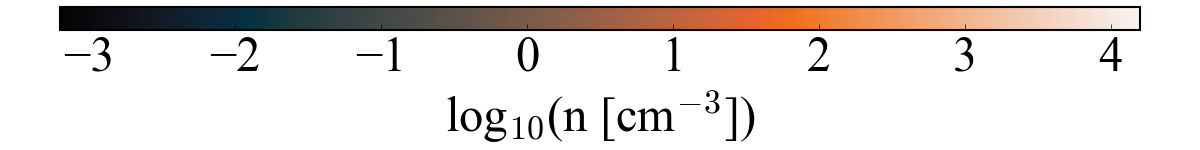}
	\includegraphics[height=0.15\textheight]{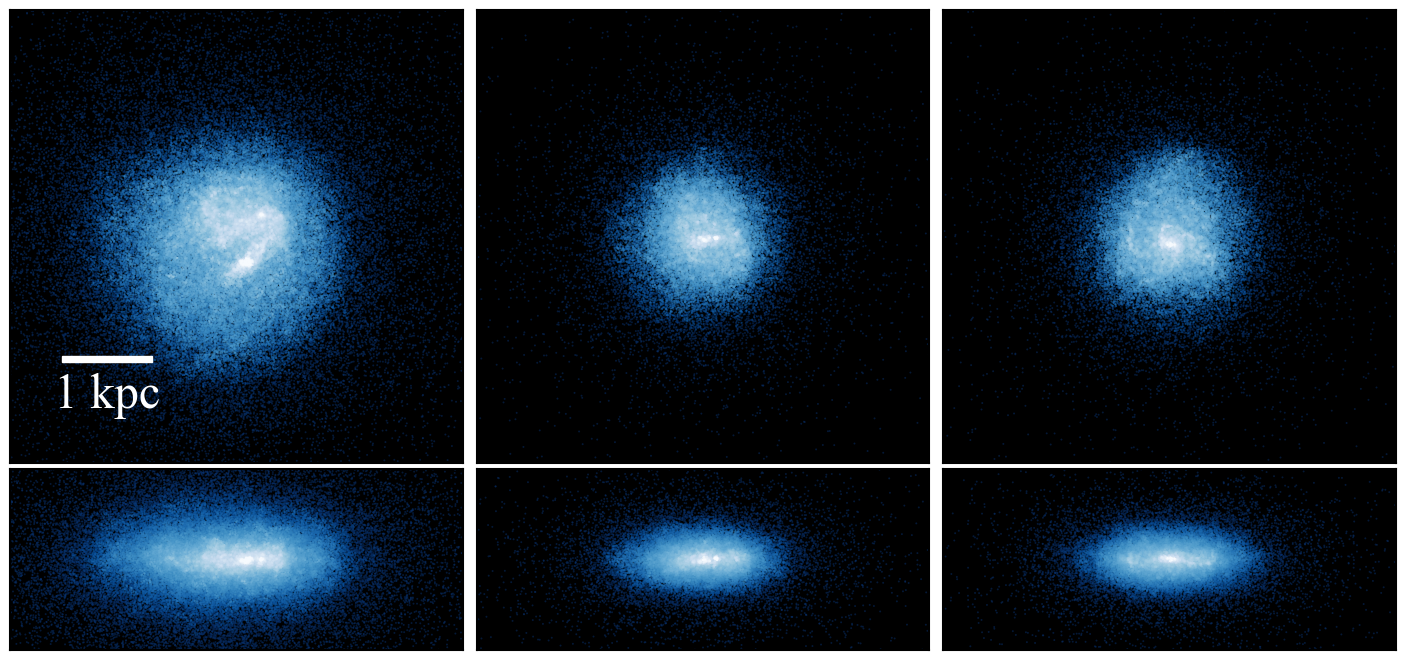}
	\includegraphics[width=1\columnwidth]{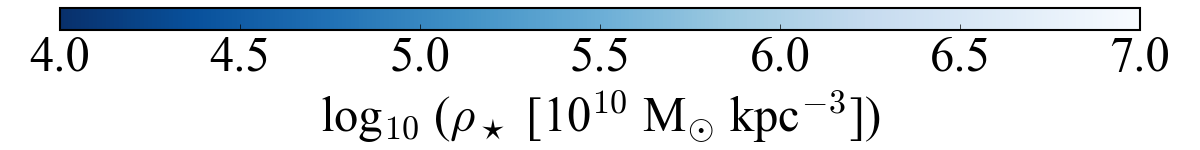}
	\caption[Galactic images for the high-resolution set at $z=3$. The gas (upper rows) and stellar (lower rows) distributions are colour coded by their respective densities]
	{As in Fig.~\ref{fig:Results_Images_low} but
	for the high-resolution simulations. 
    }
	\label{fig:Results_Images}
\end{figure}

False-colour images of the simulated galaxies at $z=3$ are displayed in Figs.~\ref{fig:Results_Images_low} and ~\ref{fig:Results_Images} for the low- and high-resolution sets, respectively.
In all cases, the reference system has been rotated based on the angular momentum of the stars to present the face-on and edge-on projections.
The images in the top rows (orange tones) show the maximum gas density along each line of sight while those in the bottom rows (blue tones) show the projected positions of the
individual stellar particles (colour coded according to their local density determined with
a spherical cubic-spline kernel).
Generally, the galaxies assume the shape of oblate ellipsoids.
The gas distribution is more extended than the stellar one and shows many filamentary features in the outskirts
resulting either from interactions with satellite systems or large-scale flows.
Most of the SN-only models (S9T, S10T and E9S) present dense gaseous and stellar concentrations at their centres, while those accounting for stellar winds are more uniform
and flattened (disc like). 

\subsection{Stellar mass versus halo mass}
\label{sec:comp_vs_obs.}

\begin{figure}
	\centering
	\includegraphics[width=1\columnwidth]{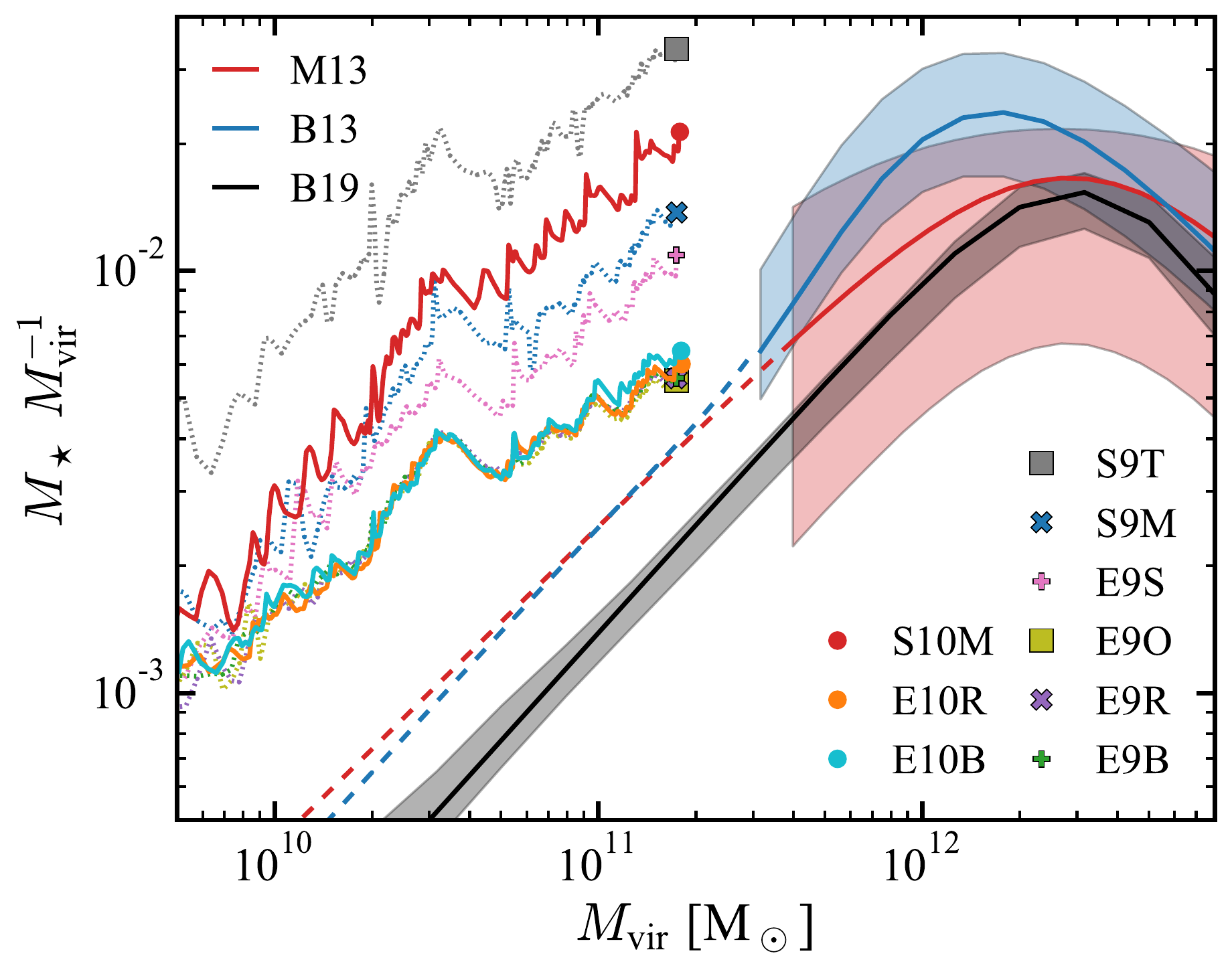}
	\caption[Stellar mass -- halo mass relation at $z=3$.]
	{The trajectories of the simulated galaxies in the $M\id{vir}$-$M_*$ plane during the redshift
	interval $9>z\geq 3$ (wiggly lines with the end point indicated by a solid symbol) are compared with the
	SMHMR at $z=3$ as determined by \citetalias{Behroozi2013}, \citetalias{Moster2013} and \citetalias{Behroozi2019}.
	Note that
	all trajectories extracted from simulations that account for stellar winds overlap almost perfectly.
    }
	\label{fig:Results_SMHM}
\end{figure}

The stellar mass - halo mass relation (SMHMR) provides a convenient way to test
whether our simulations are compatible with observations or not.
This semi-empirical relation is obtained by matching DM haloes from $N$-body simulations to estimates for the stellar mass of galaxies in a catalogue
under the assumption that $M_*$ scales monotonically with the halo mass.
In Fig.~\ref{fig:Results_SMHM}, we show SMHMRs obtained at $z=3$ 
by different authors
(smooth solid lines, \citealt{Behroozi2013}, hereafter \citetalias{Behroozi2013}; \citealt{Moster2013}, hereafter \citetalias{Moster2013}; \citealt{Behroozi2019}, hereafter \citetalias{Behroozi2019})
together with  their uncertainty (shaded regions).
These estimates consistently show that the conversion of baryons into stars is most efficient within haloes with a mass of $\sim10^{12}$ M$_\odot$ where $M_*/M\id{vir}$ is of the order of a per cent. 
In order to compare these results with our simulations we extrapolate the fits in the literature to lower halo masses (dashed lines) for which data are not available.
The wiggly lines in the left-hand side of Fig.~\ref{fig:Results_SMHM} show the trajectories
of our simulated galaxies in the $M\id{vir}$-$M_*$ plane
within the redshift range $9>z\geq 3$.
Time runs from left to right and the coloured circles highlight the end point at $z=3$.
All models (with the exception of S9T) are very close in the $M\id{vir}$-$M_*$ plane at $z=9$ but they separate more and more with time.
A rather sharp decrease in the stellar-to-halo mass ratio is caused at $z \approx 5.2$ 
by a major-merger event which increases the halo mass.
By comparing the position of the symbols to the fits from the literature,
we conclude that the SN-only models are unable to
efficiently regulate their SF via feedback and 
produce too many stars.
On the other hand, including winds brings the simulations in rather good
agreement with the SMHMR
(within the statistical uncertainties).
This holds true for both the low- and high-resolution runs and suggests that
our numerical set up provides converged global properties
for the simulated galaxies.
At earlier times (and thus for lower halo masses), the simulated galaxies are systematically shifted towards higher stellar masses with respect to the extrapolated SMHMR at $z=3$. However, this is fully consistent with estimates of the
SMHMR at higher redshifts \citepalias[see, e.g. fig.~9 in][]{Behroozi2019}.


\subsection{Stellar profiles}
\label{sec:gas_str_profiles}

\begin{figure}
	\centering
	\includegraphics[width=1\columnwidth]{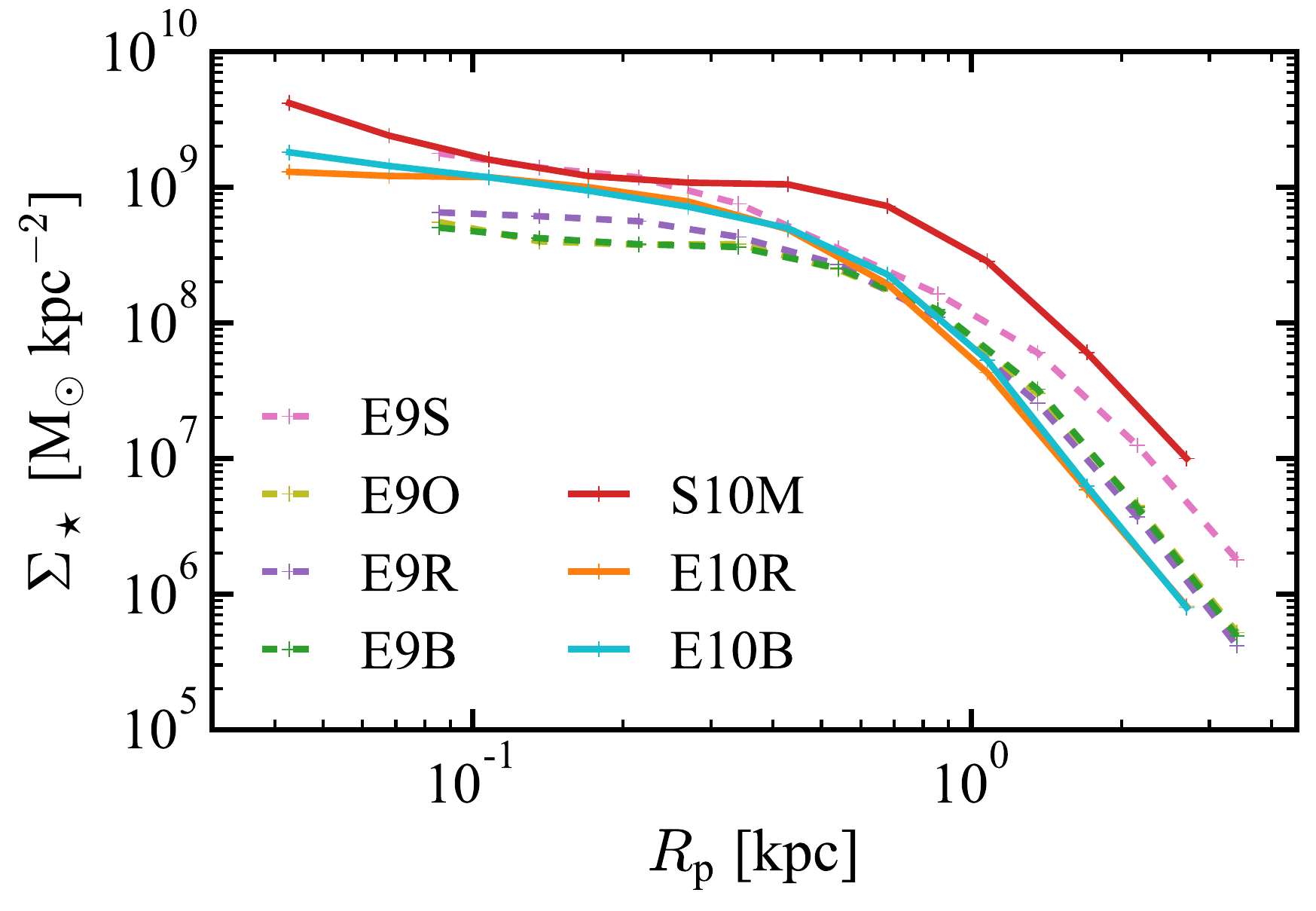}
	\caption[Stellar surface density at $z=3$]
	{Face-on stellar surface-density profiles 
	for the high-resolution and selected low-resolution runs at $z=3$.
	}
	\label{fig:Results_Surface_density_Stars}
\end{figure}

Early stellar feedback plays a key role in shaping 
galaxies as it maintains gas hot and creates pressure support which
prevents low-angular-momentum material from reaching the central regions \citep{Stinson2013}. 
The net effect is that smaller bulges are assembled.
The (face-on) projected stellar surface-density profiles of the three galaxies simulated at high resolution (Fig.~\ref{fig:Results_Surface_density_Stars})
show that S10M is denser than E10R and E10B at all radii, while the 
runs including winds present nearly identical profiles.
Towards the galactic centre, S10M shows a conspicuous increase in the stellar density
while the E-simulations exhibit a nearly constant-density
core. This difference is even more marked in the outskirts of the galaxy
where S10M is $\sim10$ times denser. 
It is worth mentioning that the profiles rapidly decline beyond 1 kpc thus
indicating that the stellar distributions are much more compact than $R\id{gal}$.
In order to distinguish between the impact of the time-dependent SN feedback and of the wind feedback, in Fig.~\ref{fig:Results_Surface_density_Stars}, we also display the results for the E9 simulations. 
Note that the run with the time-dependent SN feedback and no winds (E9S)
produces a denser stellar profile (similar to S10M) compared to
the galaxies simulated accounting for stellar winds (E9O, E9R and E9B).

\begin{figure}
	\centering
	\includegraphics[width=1\columnwidth]{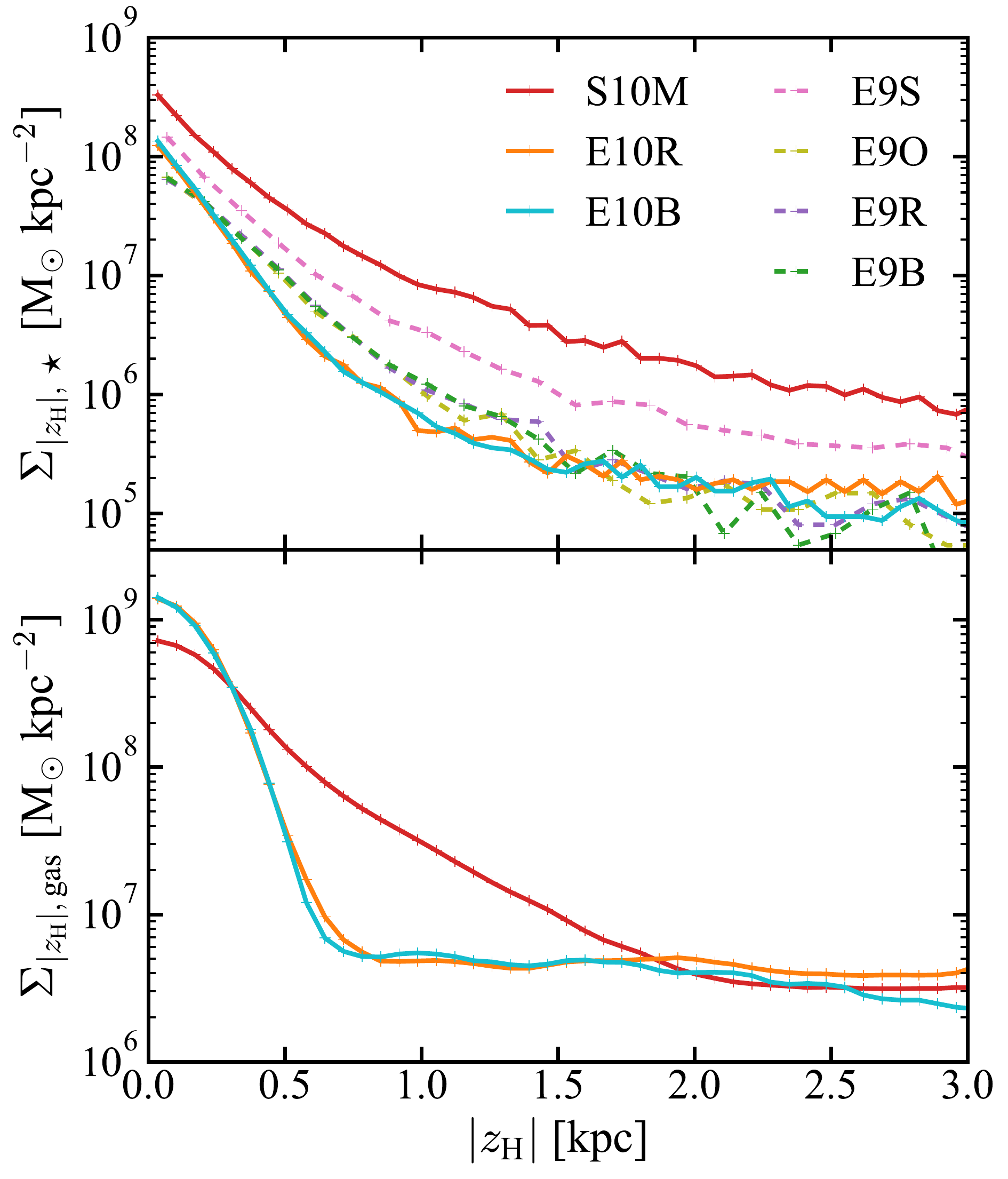}
    \caption[Vertical profiles of the gas and stars]
    {Edge-on surface-density profiles for the stars (top panel) and gas (bottom panel) in
    the high-resolution and selected low-resolution simulations at $z=3$.
    }
    \label{fig:Results_vertical_profiles}
\end{figure}

The energy injection from winds and SN into the ISM is isotropic implying that, in a disc galaxy,
feedback plays an important role in moulding the vertical surface-density profiles
of the stars.  
This quantity is analysed in the top panel of Fig.~\ref{fig:Results_vertical_profiles}
(where we use a rectangular window with transverse size $2R_\textrm{gal}$). 
Apart from the different overall normalization,
the stellar distribution in the S10M model is less flattened than in the E-counterparts
(see also Fig.~\ref{fig:Results_Images}).
The profile extracted from the E9S run lies in between those obtained from the other E9 and the S10M simulations, so the continuous SNe injection only  partially accounts for the drop in the stellar density.
Keeping this in mind, we will only consider the high-resolution runs in the remainder of this paper.

\begin{figure}
	\centering
    \includegraphics[width=1\columnwidth]{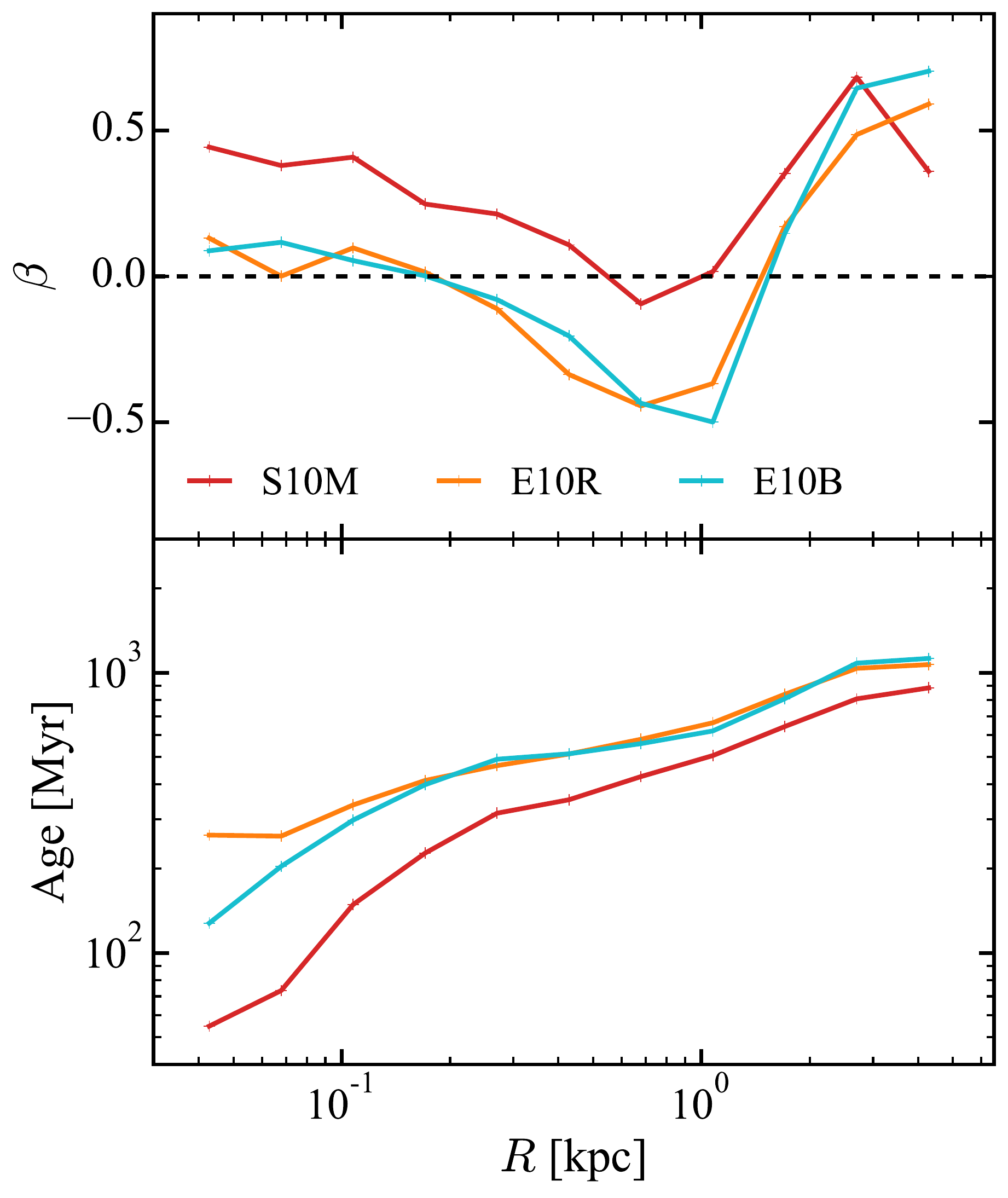}
	\caption[Spherical stellar profiles at $z=3$]
	{Spherical radial profiles of the 
	velocity anisotropy parameter (top panel) and 
	of the mass-weighted stellar age (bottom panel) at $z=3$. 
	}
	\label{fig:Results_spherical_profiles_star}
\end{figure}

In order to measure the kinematic state of the stellar orbits,
we consider the velocity anisotropy parameter, 
\begin{equation}
\label{eq:beta-paramter}
   \beta=1-\frac{\sigma^2\id{Tan}}{2\,\sigma^2\id{Rad}} \,, 
\end{equation}
where $\sigma\id{Tan}$
and $\sigma\id{Rad}$ 
denote the tangential and radial velocity dispersions, respectively. 
If all orbits in a system are purely radial passing through the galactic centre,
then $\beta=1$. If, instead, they are all circular, then $\beta \to -\infty$.
A system with an isotropic velocity distribution has $\beta=0$.
In the top panel of Fig.~\ref{fig:Results_spherical_profiles_star},
we show the radial profiles of $\beta$ for our high-resolution runs.
The shapes of the profiles look very similar in the different simulations
but $\beta$ is more biased towards radial orbits in the S10M run, at least within the central 2 kpc.
The two E-simulations have almost identical profiles which are
nearly isotropic in the centre ($R<0.2$ kpc), show a predominance of circular motion around $R=1$ kpc and are radially biased in the outskirts.
The stellar kinematic in the E10R and E10B galaxies 
thus reveals the presence of a more pronounced disc-like structure than in S10M.
On the other hand, all models present 
a tenuous stellar halo which is also discernible in the $\Sigma_*$-profiles for $R>1$ kpc.

\begin{figure}
	\centering
	\includegraphics[width=1\columnwidth]{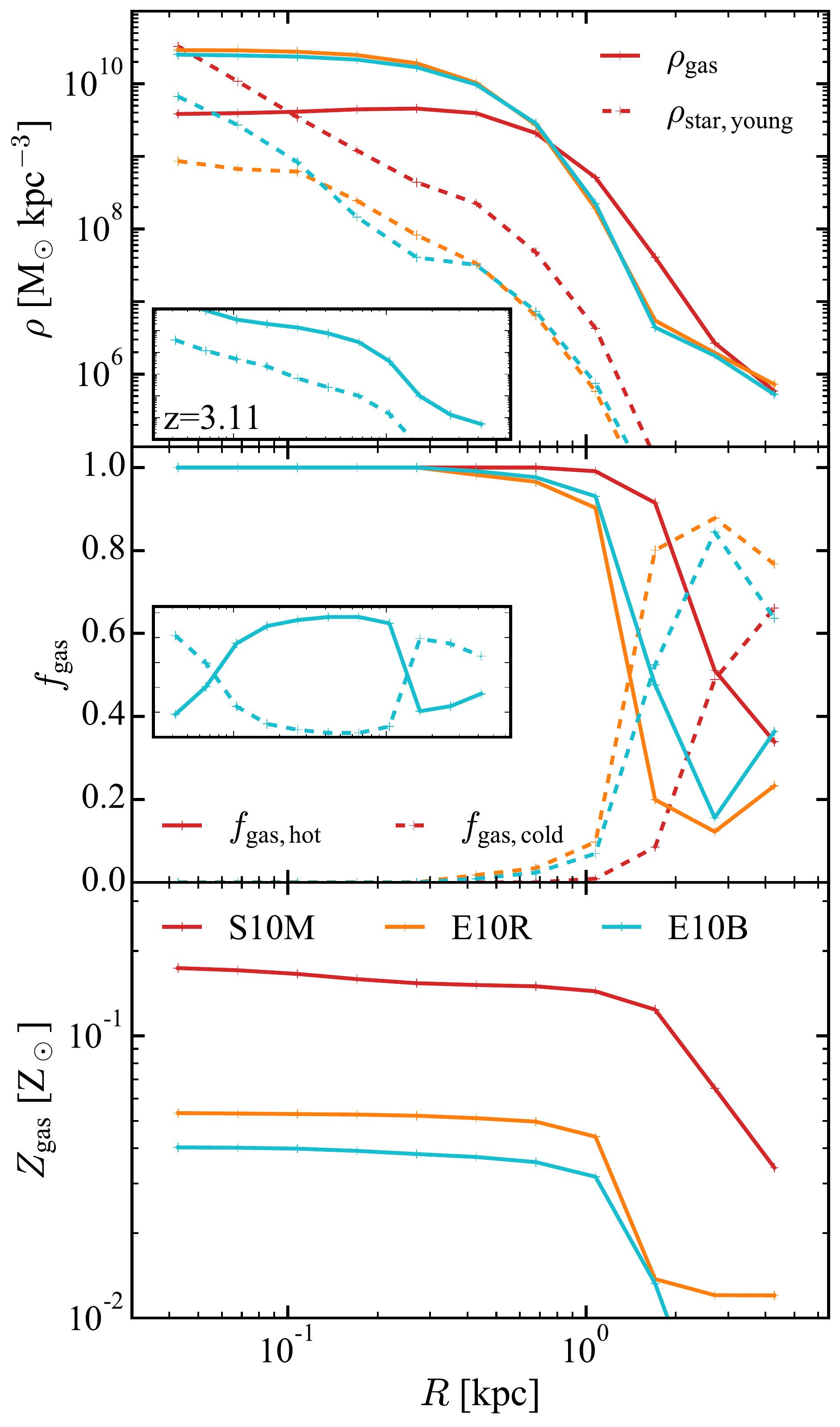}
	\caption[Spherical gas profiles at $z=3$]
	{Spherical radial profiles for different galaxy properties at $z=3$. 
	The top panel shows the mass-density profiles for the gas (solid) and for the recently formed stars ($t_*<\SI{50}{\Myr}$, dashed). The middle panel reveals
	the radial dependence of the fractions of cold and hot gas. Finally, the bottom panel displays the average metallicity of the gas. The insets in the top two panels refer to the E10B galaxy at $z=3.11$ and have compressed axes that represent the same
	range as the extended panels.
	}
	\label{fig:Results_spherical_profiles_gas}
\end{figure}

The radial mass-weighted age profiles of the stellar populations at $z=3$ are presented in
the bottom panel of Fig.~\ref{fig:Results_spherical_profiles_star}.
For all galaxies, the mean stellar age increases with $R$ but
it turns out that the stars in S10M are on average younger than in E10R and E10B at all radii. The difference is more pronounced in the central regions ($R<\SI{0.2}{\kpc}$)
which have been actively forming stars shortly before $z=3$ (see also Fig.~\ref{fig:Results_spherical_profiles_gas}).
This reflects the higher efficiency of early stellar feedback in regulating local
star formation.

\subsection{Gas profiles}
\label{sec:gas-profiles}

It is interesting to investigate how the gas and stellar distributions in the simulated galaxies relate to each other. 
In the bottom panel of Fig.~\ref{fig:Results_vertical_profiles}, we show the
edge-on surface-density profiles of the gas at $z=3$. The gas distribution in the E10R and E10B  galaxies is more concentrated than in S10M and reaches higher densities
 at the centre (which compensates for the lower stellar content). The gas profiles of the E-models, however,
show an abrupt decline at $|z\id{H}|\approx 0.6$~kpc which can be used to define the edge of the galaxies. 
Additionally, this decline marks a clear transition between a disc-like structure and its surrounding halo, which is not present in the S10M galaxy.
We conclude that the E10R and E10B galaxies present
a flatter, more disc-like gas distribution than S10M (see also the top panels in Fig.~\ref{fig:Results_Images}).

The top panel of Fig.~\ref{fig:Results_spherical_profiles_gas}
presents the spherical gas-density profiles (solid lines) for the galaxies
at $z=3$ while the middle panel shows the mass fractions of `cold' ($T<2\times10^4$ K, from which stars form in the simulations) 
and hot ($T>2\times10^4$ K) gas as a function of the distance from the galaxy centre.
Remarkably, the E-simulations contain ten times more gas at their centres than the S10M model. In all cases, the gas density stays nearly constant
for $R\lesssim\SI{0.6}{\kpc}$ and drops rapidly at larger distances.
All the gas in this extended region with uniform density is hot due to the presence of
young stars which have recently injected energy into the ISM. Cold gas is only present in the outer regions
where little or no star formation took place in the recent past.
The galaxies at $z=3$ happen to be in a transient sterilised state: feedback from a recent star-formation episode prevents new star formation (see also Section~\ref{sec:outflows}). 
In order to demonstrate that this is indeed the case, in the insets of the top and middle panels we show again the profiles for the E10B galaxy but this time we evaluate them at $z=3.11$ (i.e. approximately 80 Myr before $z=3$): the presence of cold and dense gas in the central regions now makes star formation possible.

The bottom panel of Fig.~\ref{fig:Results_spherical_profiles_gas} shows the metallicity profiles of the ISM. As already seen in Table~\ref{tab:gal_properties},
the S10M galaxy is substantially more metal rich than the E-runs (by a factor 2.9 and 4 with respect to E10R and E10B, respectively) but the shape of the metal
distribution is similar in all objects with an extended uniform core and 
a rapidly declining profile for $R\gtrsim\SI{1.5}{\kpc}$. 
This drop reflects the finite size of the stellar components, the feedback 
efficiency in expelling metals, and
the infall of metal-poor gas from the circumgalactic medium.
The lower metal content of the E-galaxies makes gas cooling less efficient 
and thus halts the formation of new stars for longer times after a burst of star
formation takes place. The different normalization of the metal profiles in E10R and E10B
is a direct consequence of the yields shown in Fig.~\ref{fig:Methods_final}.

\begin{figure}
	\centering
	\includegraphics[width=1\columnwidth]{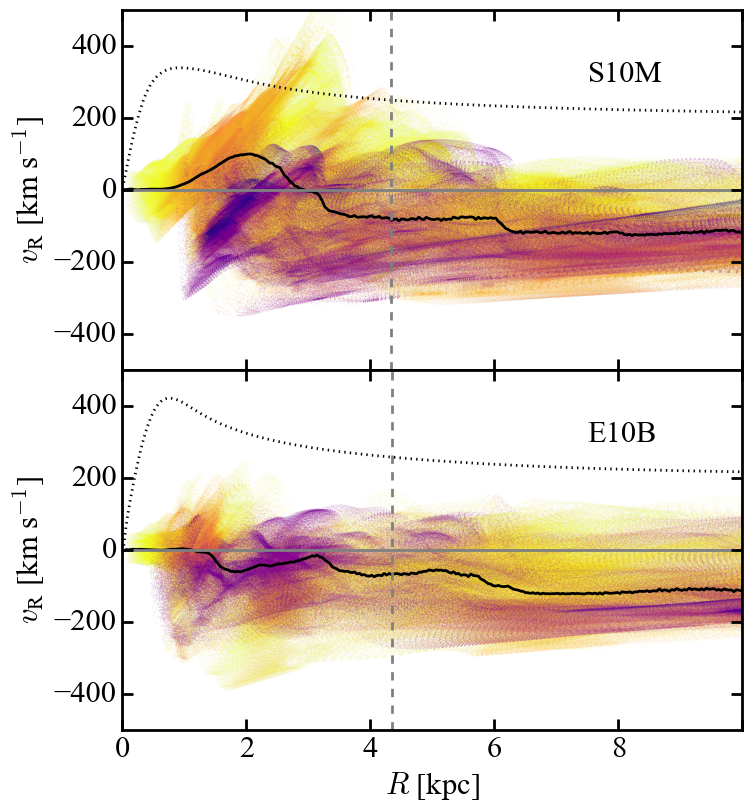}	
	\includegraphics[width=0.9\columnwidth]{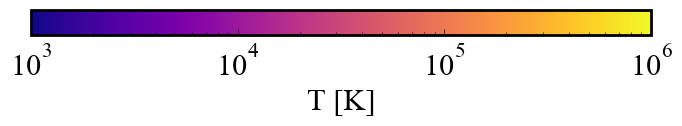} \hspace{-1.3cm}
	\caption[Phasespace]
	{Radial phase-space diagrams 
	for the gas in the S10M (top panel) and E10B (bottom panel) simulations at $z=3$.
	The colour coding reflects the gas temperature while
	the solid and dotted black lines represent the 
	mass-weighted mean
	radial velocity and the escape velocity, respectively.
	The vertical dashed grey line indicates the galaxy radius $R\id{gal}$. 
	}
	\label{fig:Results_Phasespace}
\end{figure}
\subsection{Outflows and metal enrichment}
\label{sec:outflows}
In Fig.~\ref{fig:Results_Phasespace},
we present radial phase-space diagrams of the gas component at $z=3$
colour-coded by temperature.
Outflows of hot gas with radial velocities exceeding the escape velocity (dotted line)
are clearly noticeable in the S10M galaxy (top panel).
The (mass-weighted) mean gas velocity (solid line) is positive for $R<3.0$ kpc (with a peak value of 100 km s$^{-1}$)
and negative at larger radii implying
that there is a net inflow of material within 
$R\id{gal}$ (vertical dashed line).
The E10B galaxy (bottom panel) shows a similar net inflow of material but does not
present fast outflowing gas that could escape the system.

\begin{figure}
	\centering
	\includegraphics[width=1\columnwidth]{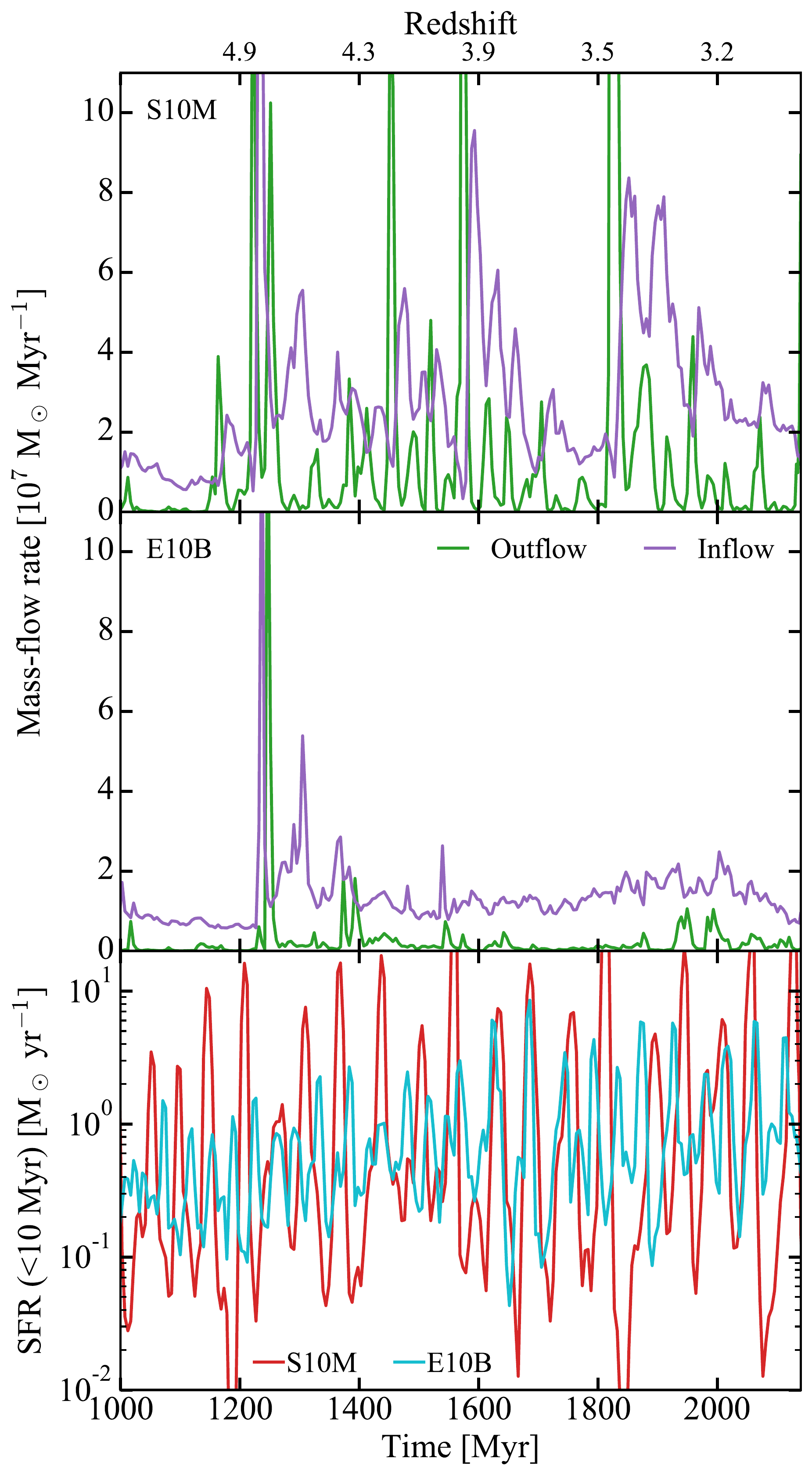}
	\caption[Flow rates]
	{The inward and outward gas-mass flow rates measured at $R=2$ kpc in the S10M (top panel) 
	and E10B (middle panel) galaxies are compared with the corresponding
	SFR averaged over \SI{10}{\Myr} (bottom panel). 
	}
	\label{fig:Results_Outflows}
\end{figure}

In order to better understand how early and SN feedback influence galactic outflows
and push the gas into the circumgalactic medium,
in Fig.~\ref{fig:Results_Outflows}, we compare 
the inward and outward mass-flow rates calculated at $R=\SI{2}{\kpc}$
with the SFR of the simulated galaxies (averaged over a time-span of \SI{10}{\Myr}).
The first thing to notice is that star formation is bursty in both galaxies (as expected from a self-regulating process) 
but the amplitude of the variations is much larger in S10M -- the coefficient of variation $CV$ (i.e. the ratio of the standard deviation of the SFR to the mean) is twice as large than in E10B (namely, $CV_\mathrm{S10M}=2.6$ while $CV_\mathrm{E10B}=1.3$). The highest peaks of star
formation trigger gas outflows that go past \SI{2}{\kpc} (like the one at $z\simeq 4.8$ following a major galaxy merger). These outflow maxima are followed by 
sudden increases of the inflow rates, suggesting that a galactic fountain is in place.
These features are much more prominent in S10M than in E10B. 

All this is consistent with the picture in which 
stellar winds heat up the ISM and locally inhibit further star formation thus leading to a more uniform star-formation history.
The reduced (and less spatially clustered) SN activity is then unable to launch high-speed
galactic winds. This is in line with the recent
finding that early stellar feedback suppresses galactic outflows in galaxies hosted by lower-mass haloes than those considered here \citep{Smith2021}.

\begin{figure}
    \centering
    \includegraphics[width=1\columnwidth]{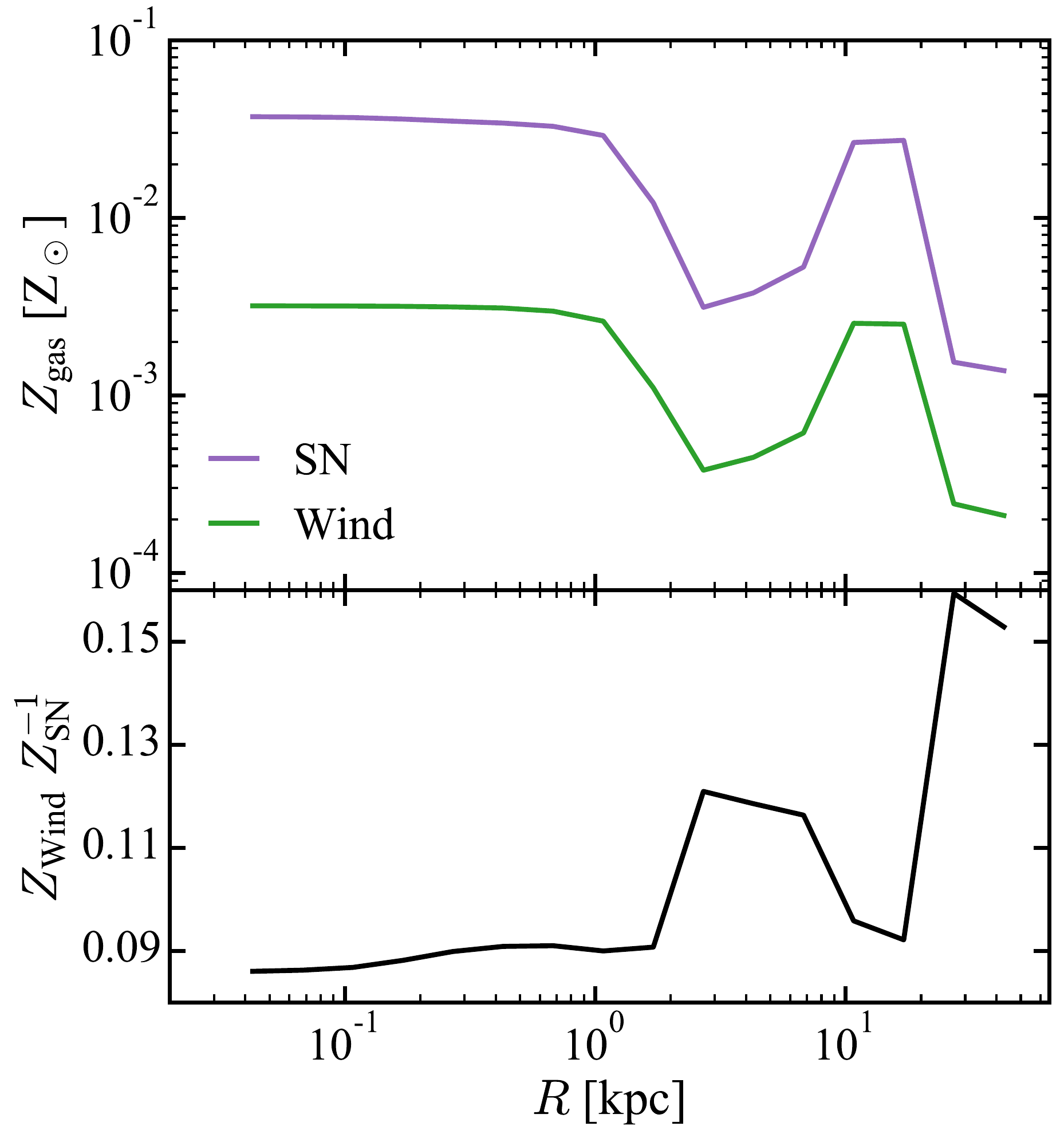}
    \caption[Profiles of metallicity ratio]
    {Top panel: spherical radial profiles of the mass-weighted gas metallicity (for the E10B galaxy at $z=3$) obtained by separating the contributions from the metals entrained in stellar winds and from those ejected by SNe. Bottom panel: ratio between the profiles shown in the top panel.  
    }
    \label{fig:Results_metallicity_prof}
\end{figure}

\begin{figure}
    \centering
    \includegraphics[width=0.49\columnwidth]{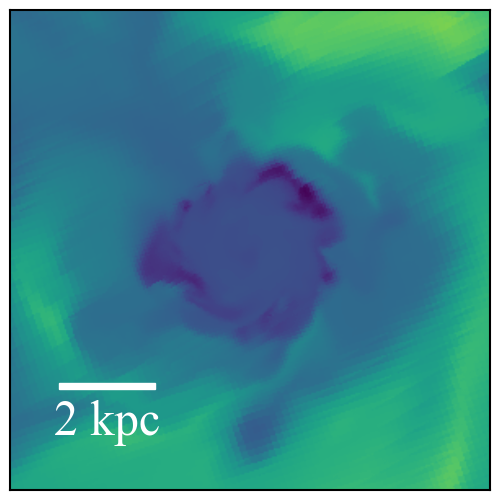}
    \includegraphics[width=0.49\columnwidth]{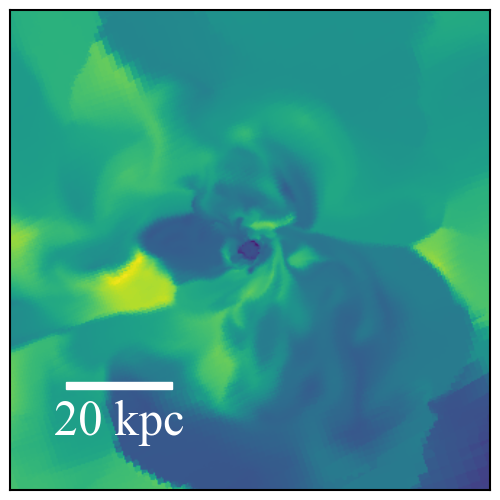}
    \includegraphics[width=1\columnwidth]{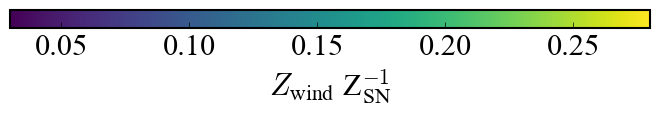}
    \caption[Slice of metallicity ratio]
    {Face-on maps of the ratio between the wind- and SN-generated metallicities
    for the E10B galaxy (left-hand panel) and its halo (right-hand panel) at $z=3$. 
    }
    \label{fig:Results_metallicity_image}
\end{figure}

It is worth noticing that also the mass-inflow rates at $R=2$ kpc are lower in E10B than in S10M. This could be for two reasons: (i) since outflows are weaker,
there is less material that could fall back and/or
(ii) the infall of recycled and pristine gas on to the galaxies is prevented by early feedback which keeps the gas hot.

Finally, we investigate if metals ejected by stellar winds follow a different spatial distribution than those ejected by SNe. 
In a recent theoretical study aiming at explaining the large observed scatter of
N/O at low O/H,
\citet{Roy2021} conjectured that wind metals are more likely to remain locked up within a low-metallicity dwarf galaxy than SN metals. 
We test whether this scenario takes place in our simulated galaxies
which, however, are hosted by substantially more massive haloes than those analysed
in \citet{Roy2021}.
The spherical radial profiles for the E10B galaxy at $z=3$ shown in 
Fig.~\ref{fig:Results_metallicity_prof} 
reveal that
nearly 90 per cent of the metals which are found within $R\id{gal}$ (and 80 per cent of those at $R\id{vir}$)  have been ejected by SNe. The mean metallicities generated by winds and SNe stay basically constant
in the innermost regions (where also the stars are found) and suddenly drop by an order of magnitude at $R\simeq\SI{1.5}{\kpc}$
revealing that the circumgalactic medium is relatively metal poor 
(the sharp peak
around $R=\SI{10}{\kpc}$ is due to a satellite galaxy).
Maps of the relative distribution of metals due to winds and SNe  (Fig.~\ref{fig:Results_metallicity_image}) confirm that the two types of metals
are very well mixed within the galaxy. This reflects the fact that
the material emitted by winds does not travel very far from the massive stars
and is subsequently swept up by the faster SN ejecta. 
Beyond the galaxy, the maps are more complex due to the interplay of gas accretion, outflows, and the presence of satellite galaxies (passing by or being ripped appart).
Apart from a few localized features, a general trend is noticeable: the relative
importance of wind metals slightly increases with $R$, probably due to the  $Z$ dependence of the wind-metal ejecta (see Fig.~\ref{fig:Methods_final}).



\section{Summary}
\label{sec:Summary}

Main-sequence and post-main-sequence winds from massive stars 
provide a continuous injection of kinetic energy, mass, and metals into the ISM.
In order to quantify the corresponding yields, 
we compute different sets of evolutionary tracks using the \textsc{Mesa} code
and accounting for the presence of binary systems and of a metallicity-dependent distribution of initial rotational velocities. We find that:
\begin{enumerate}
\item 
For the most-massive models,
the ratio between the
kinetic-energy yields of stellar winds and SNe
ranges from a few per cent to more than a hundred per cent depending on the input
parameters ($Z, \upsilon_\mathrm{ini}$, binarity).
Crucially, this energy becomes available 
on time-scales shorter than the free-fall time of a young cluster.
\item A stellar population consisting of rotating and binary stars generates substantially stronger mechanical feedback compared to standard models based on single non-rotating stars, especially at low metallicity.
In fact, both binaries and rotation significantly flatten the otherwise steep metallicity dependence of the mechanical-energy yield (see Section~\ref{sec:fdbk_stel-pop}).
Additionally, the mass and metal yields are also enhanced.
\end{enumerate}

As a second step, we implement the feedback yields derived from our stellar evolutionary models into \textsc{Ramses} and
run a suite of simulations which follow the formation and evolution of a galaxy until redshift $3$ (at which we achieve a nominal spatial resolution of \SI{34}{\pc}).
We follow the central galaxy hosted by a DM halo with a final mass of $1.8 \times 10^{11}$ M$_\odot$ and
use the same IC to compare the galaxies generated with different versions of our
early feedback scheme (E models) with the galaxy obtained considering SN-only feedback in the standard \textsc{Ramses} implementation (S model).
For the E models, we consider three different options:
non-rotating single stars, rotating single stars, and a mix of rotating single stars and binary stars. 

It is important to stress that modelling stellar mass-loss requires a number of assumptions and even our detailed evolutionary models bear large uncertainties. The complex interaction of the wind ejecta with the circumstellar material introduces further uncertainties in the calculations (see the discussion in Section~\ref{sec:fdbk_discussion}).
In the absence of a consensus in the literature regarding the fraction of emitted energy which is dissipated into heat on sub-grid scales, we inject the whole energy released by the winds and SNe into the ISM and solve the equations of fluid dynamics to determine the gas flows on spatially resolved scales. In this sense, our investigation quantifies the maximum effect that could be possibly driven by stellar winds on galaxy formation.
Our main findings are as follows:
\begin{enumerate}\setcounter{enumi}{2}
    \item In the E models, the stellar mass is reduced by a factor of three compared with the S model. This makes sure that the stellar-mass-to-halo-mass ratio
    is consistent with current semi-empirical estimates (see Fig.~\ref{fig:Results_SMHM}).
    \item The stellar surface-density profile of the E galaxies flattens in the central region, contrary to the outcome of the S model which shows a central cusp  
    (see Fig.~\ref{fig:Results_Surface_density_Stars}). Additionally, the stars in the E galaxies have a lower
    anisotropy parameter, indicating that the structures are more rotationally supported
    (see Fig.~\ref{fig:Results_spherical_profiles_star}). 
    \item 
    Accounting for wind feedback leads to a smoother and less bursty SFR, less strong outflows and even reduced accretion flows.
    \item 
    All the E galaxies have very similar stellar and gas masses. However,
    those including binary stars turn out to be more metal poor than those with single stars (while no important difference is noticeable between rotating and non-rotating models). A caveat to this is that we neglect the impact of binarity and rotation on the nucleosynthesis of SNe.
    \item The final spatial distribution of the metals which have been
    entrained in stellar winds or ejected by SNe is very similar within the galaxy and also in the circumgalactic medium. SN metals account for nearly 90 per cent of the total with a slight decrease in the outermost regions.
\end{enumerate}


\section*{Acknowledgements}
We would like to thank R. Teyssier for making the \textsc{Ramses} code publicly available. We also thank J. Mackey and S. Geen for valuable discussions. 
Some of the figures were produced using the \textsc{yt} package \citep{Turk2011}. This research has made use of the VizieR catalogue access tool, CDS, Strasbourg, France \citep{Ochsenbein2000}. The original description of the VizieR service was published in 2000, A\&AS 143, 23.
This work was carried out within the Collaborative Research Centre 956, sub-project C04, funded by the Deutsche Forschungsgemeinschaft (DFG) – project ID 184018867.
We gratefully acknowledge the Gauss Centre for Supercomputing e.V. (\url{www.gauss-centre.eu}) for funding this project by providing computing time on the GCS Supercomputer SuperMUC-NG at Leibniz Supercomputing Centre (\url{www.lrz.de}). 
We acknowledge the Max-Planck-Society for providing computing time on the MPG Supercomputer Cobra at the Max Planck Computing and Data Facility. 
YAF is part of the International Max Planck research school in Astronomy and Astrophysics and guest at the Max-Planck-Institute for Radio Astronomy.

\section*{Data availability}
The data underlying this article will be shared on reasonable request to the corresponding author.



\typeout{} 
\bibliographystyle{mnras}
\bibliography{Quellen} 




\appendix
\section{Mass yield of the stellar models}
\label{sec:Appendix_massyield}

\begin{figure*}
	\centering
	\includegraphics[width=1\textwidth]{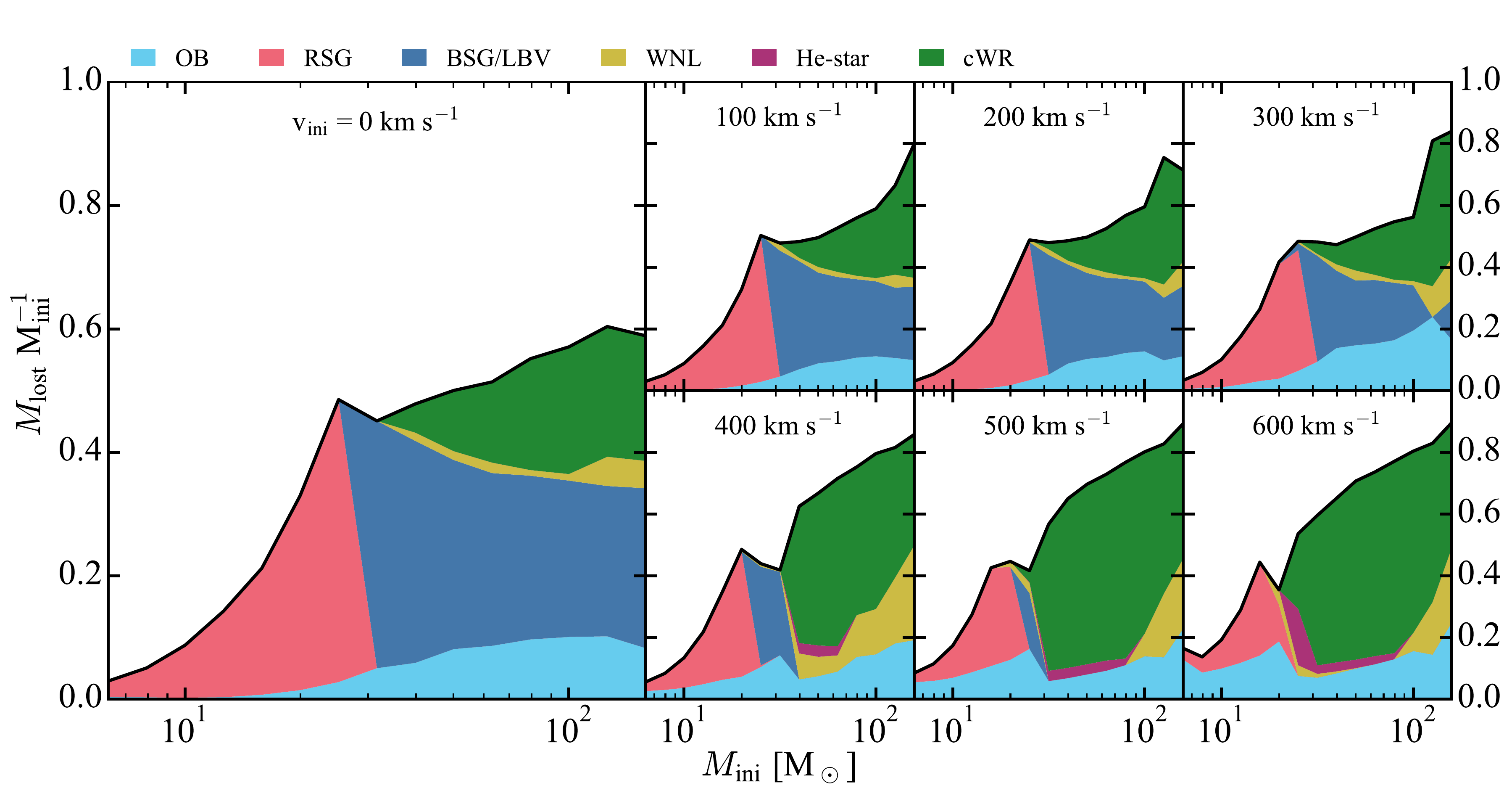}
	\caption{As in Fig.~\ref{fig:Methods_Single_Initialmass_Energy} but for the
	fractional mass-loss of single stars with $Z=0.004$.
	}
	\label{fig:Methods_Single_Initialmass_Mass}
\end{figure*}

\begin{figure*}
	\centering
	\includegraphics[width=1\textwidth]{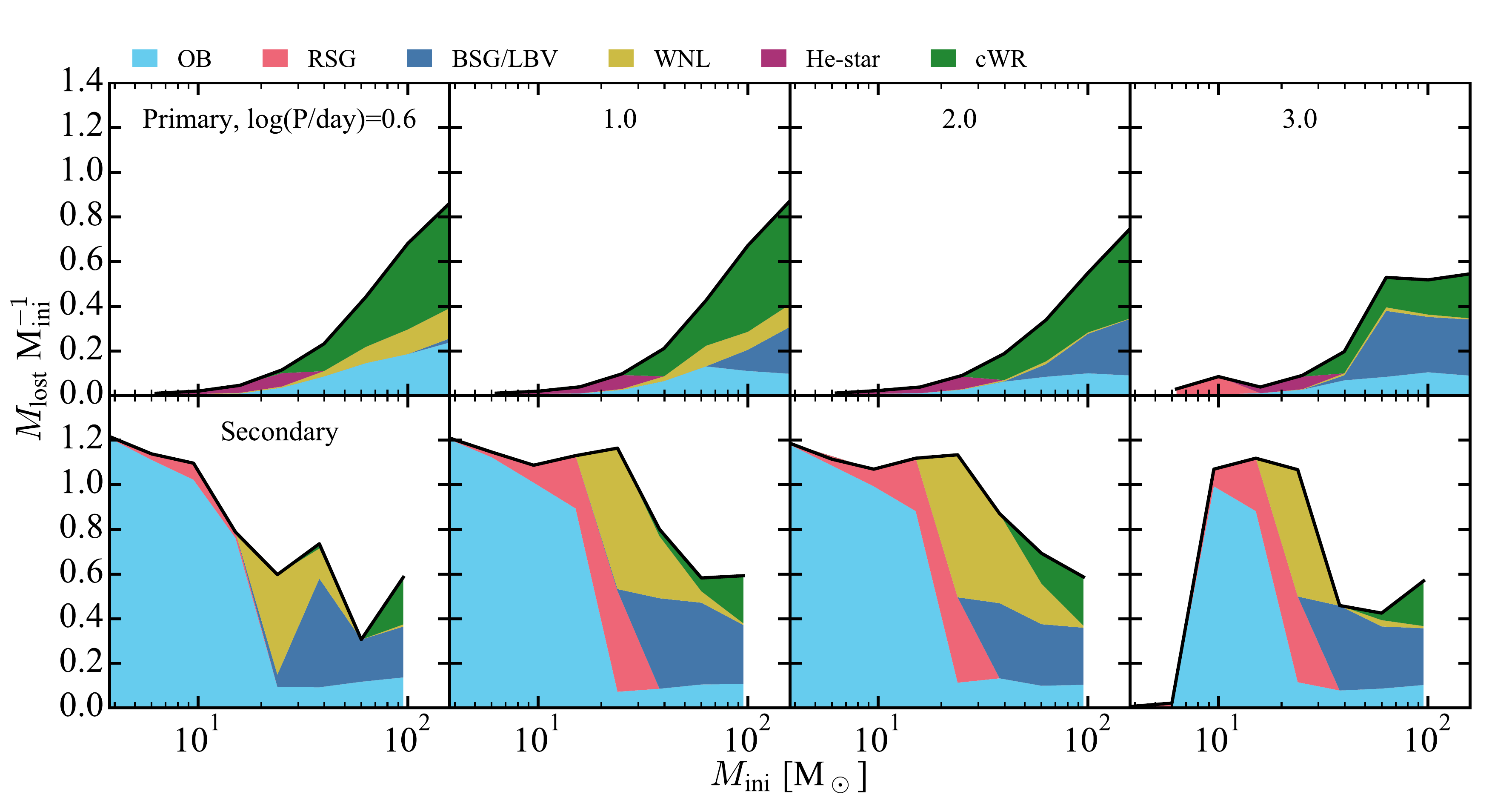}
	\caption{As in Fig.~\ref{fig:Methods_Binary_Initialmass_Energy} but for the fractional
	mass-loss of binary stars with $Z=0.004$.
	}
	\label{fig:Methods_Binary_Initialmass_Mass}
\end{figure*}

\subsection{Ejected mass of the different stellar models}

Fig.~\ref{fig:Methods_Single_Initialmass_Mass} shows the fraction of the initial mass that the single star models eject during the different phases of their lifetime. 
We can see that, even at this relatively low metallicity, our models above $\approx\M{30}$ lose between \num{50} and \num{80} per cent of their initial mass through their stellar winds, with the most massive models returning several tens of solar masses of material to the ISM on the order of some \SI{e6}{\year}. Overall, 
the fractional ejected mass increases for higher masses. The relative contribution of the OB dwarf phase to the ejected mass is lower compared to the contribution to the kinetic energy. The mass-loss of the non-rotating models below \M{25} is completely dominated by the RSG phase, in stark contrast with the wind energy (Fig.~\ref{fig:Methods_Single_Initialmass_Energy}). Instead, for higher masses, the crucial phase appears to be the BSG/LBV phase, where stellar models eject most of their outer, H-rich layers. In fact, while in the top-right corner of the HRD, near or past the so-called Humphreys-Davidson limit \citep{Humphreys1994}, the extreme proximity to the Eddington limit \citep{Sanyal2015,Grassitelli2021} and the bi-stability mechanism \citep{Vink2001}, lead to mass-loss rates as high as several \SI{e-4}{\Msun\per\year}.
For higher rotation velocities, the fraction of ejected mass increases. However, for the highest rotation velocity of \SI{600}{\kilo\meter\per\second}, our most massive stellar models do not evolve toward lower effective temperatures; hence they do not experience neither a BSG/LBV nor a RSG phase. Instead, most of the mass is lost during the WR phases.

Fig.~\ref{fig:Methods_Binary_Initialmass_Mass} shows the fractional mass-loss of our binary models. 
In the closest binary systems, the primary fills its Roche lobe and is stripped of the outer, H-rich stellar layers well before it can become a supergiant. As such, most of the mass-loss by stellar wind leaves the system during a cWR phase following the case A mass-transfer phase. However, compared to the single star models, the primary alone reaches helium exhaustion while retaining more of its initial mass, due to the lack of a RSG phase (and a reduced importance of the BSG/LBV phase) with only the evolutionary models more massive than \M{60} loosing more than half of their initial mass.
In the longest period binary models, the primary undergoes mass-transfer only once as it approaches its post-main-sequence phase, reducing the overall duration of the cWR phase and, in turn, the fractional mass-loss. None the less, compared to the single star models, the primaries appear to return less material to the ISM.
However, we can see that, due to mass-transfer, the secondary stellar model can eject even more than its initial mass (Fig.~\ref{fig:Methods_Binary_Initialmass_Mass}). This includes material which is lost during the mass-transfer phase, in which accretion to the secondary is limited by the model reaching critical rotation \citep{Langer2004,Paxton2015}. In such a situation, only a limited fraction of material lost by the primary is stably accreted, while the remaining is considered lost via the stellar wind of the accretor.
For the primaries, as the single stars, the fractional mass-loss increases towards higher mass. Instead, the lower mass secondaries have higher fractional mass-losses.
Considering that mass-transfer takes place predominantly during the secondary's main-sequence, the importance of the post-main-sequence phase is lower compared to the single stars.

\subsection{Time evolution of the cumulative ejected mass}
\begin{figure}
	\centering
	\includegraphics[width=1\columnwidth]{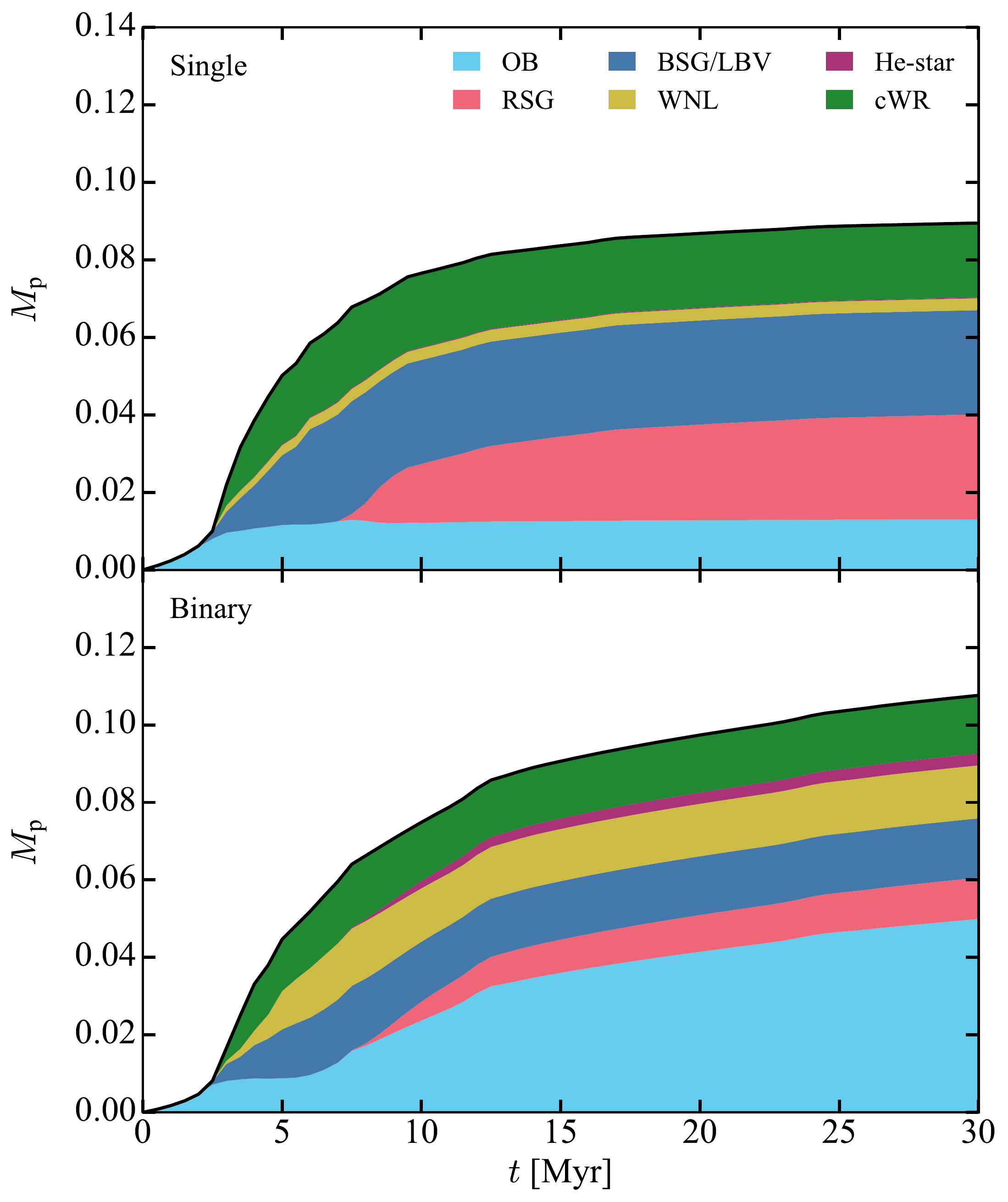}
	\caption{
	As in Fig.~\ref{fig:Methods_Time_Energy} but for the
	cumulative ejected mass in the form of winds by a coeval simple stellar population with metallicity $Z=0.004$. 
	}
	\label{fig:Methods_Time_Mass}
\end{figure}

The fractional mass-loss of a stellar population consisting of single rotating stars up to helium exhaustion is presented in the upper panel of Fig.~\ref{fig:Methods_Time_Mass}. Colours indicate the mass ejected in each phase. 
The contribution from the stellar models with lower $M_\mathrm{ini}$ to the mass budget is higher than for the energy, leading to \num{45} per cent of the total mass being ejected after the first \SI{5}{\Myr}. The cWR phase is less important, contributing only around \num{22} per cent of the total. The total mass-loss is dominated by stellar models during the BSG/LBV and the RSG phase, where the RSG phase adds mass after the first \SI{8}{\Myr} when the first lower-mass stars evolve away from the main-sequence.

In comparison with single stars, a population consisting of only binaries ejects a $\approx 30$ per cent higher total mass fraction (see lower panel of Fig.~\ref{fig:Methods_Time_Mass}), mostly due to the contribution of lower mass stars after the first few million years. The increase in the mass yield is due to non-complete accretion of material during the mass-transfer phase, which takes place while the star appears as a OB dwarf. Therefore, the OB dwarf phase has a high contribution for binaries, while the RSG and BSG/LBV phase is relatively more important for single stars.

\section{Initial rotation velocities}
\label{sec:Appendix_rotation}

\begin{figure}
	\centering
	\includegraphics[width=0.5\textwidth]{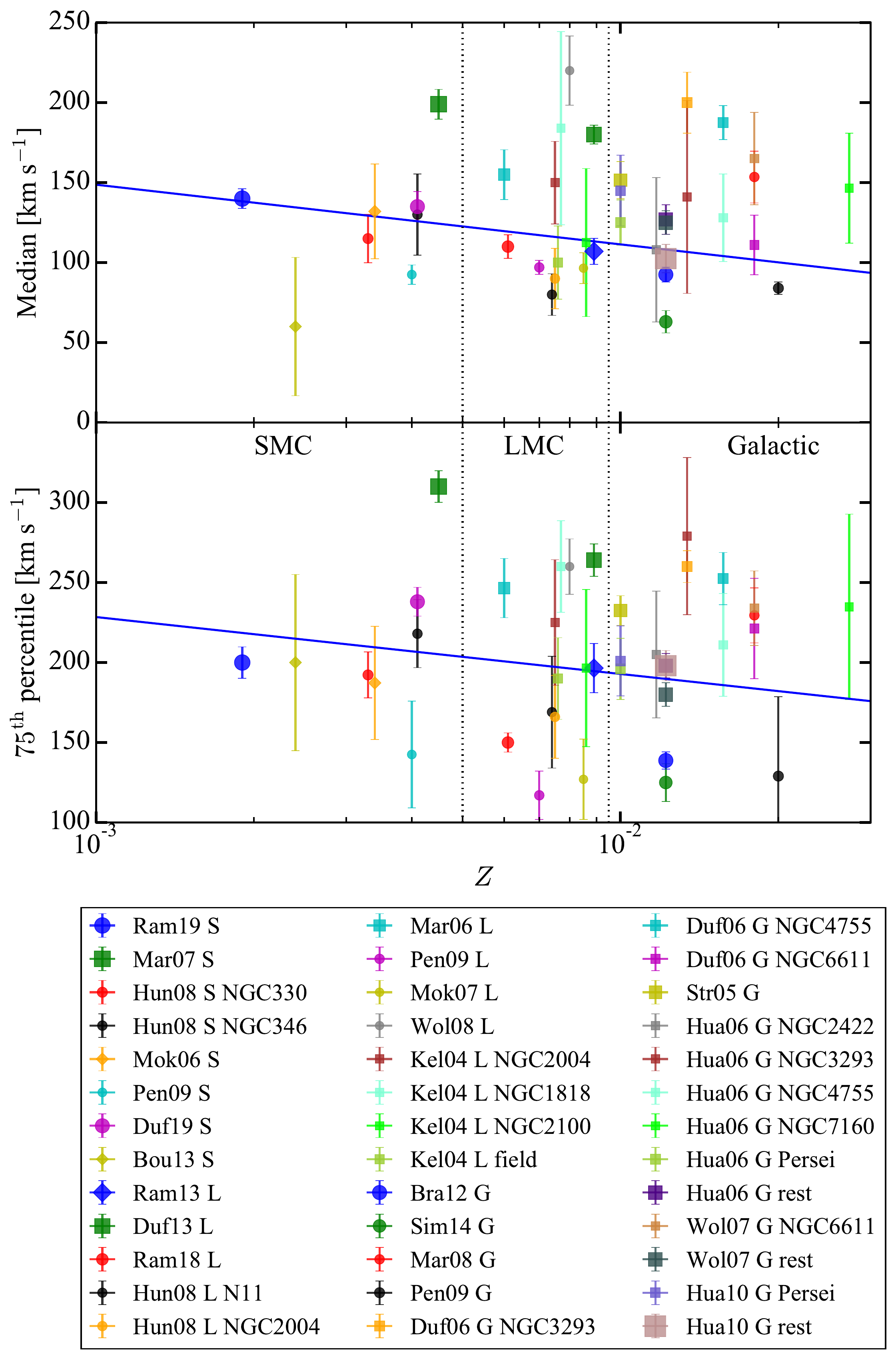}
	\caption{The median (top panel) and the 75$^\mathrm{th}$ percentile (bottom panel) of the projected rotation velocities for different stellar clusters. The size of the symbols reflects the number of stars considered in each study while the shape of the symbols indicates the type of the stars (circles, squares and diamonds for OB, B, and O stars, respectively). The blue line shows the best-fitting relation. The dotted vertical lines separate SMC, LMC, and Galactic samples.
	}
	\label{fig:Methods_fits_rot}
\end{figure}

The distribution of initial rotation velocities in a stellar population, ${\mathcal P}(\upsilon_\mathrm{ini})$, is expected
to be metallicity-dependent. 
In fact, stars with lower metallicities should 
present higher values of $\upsilon_\mathrm{ini}$ 
as the loss of angular momentum is less efficient during their formation phase
\citep[e.g. ][]{Chiappini2006}.
\citet{RamirezAgudelo2013} measured ${\mathcal P}(\upsilon_\mathrm{ini})$
for a large sample of O-type stars in the Tarantula nebula
and 
identified two components: a low-velocity peak and a high-velocity tail so that \num{20} per cent of the population consists of fast rotators with velocities above \SI{300}{\kilo\meter\per\second}.
However, the detailed form of ${\mathcal P}(\upsilon_\mathrm{ini})$
as a function of $Z$
is largely unconstrained primarily due to the limited data available.

In order to get an estimate of the metallicity dependence of 
${\mathcal P}(\upsilon_\mathrm{ini})$,
we collect estimates in the literature\footnote{We exclude any samples that are selected by the rotation velocity and any identified binary stars. Evolved stars rotate on average slower due to the loss of angular momentum via stellar wind and the significantly larger radii, therefore we focus the analysis on young open clusters. We disregard instead studies on Be-stars, as stars in such samples are selected for being near-critical rotators \citep{Martayan2006, Martayan2007,Rivinius2013}. We take the metallicity of the sample either directly from the quoted literature or, if none is provided, from other estimates.} of the projected rotation velocity of young clusters in
Galactic \citep{Strom2005, Dufton2006, HuangGies2006, Wolff2007, Martayan2008, Penny2009, Huang2010, Braganca2012, Simon-Diaz2014}, LMC \citep{Keller2004, Martayan2006, Mokiem2007, Hunter2008, Wolff2008, Penny2009, RamirezAgudelo2013, Dufton2013, Ramachandran2018} and Small Magellanic Cloud \citep[SMC,][]{Mokiem2006, Martayan2007, Hunter2008, Penny2009, Bouret2013, Ramachandran2019, Dufton2019} environments. 
In Fig.~\ref{fig:Methods_fits_rot}, we plot the resulting medians and
75$^\mathrm{th}$ percentiles (together with their standard uncertainties) as a function of $Z$.
We use a weighted least-squares method to fit these estimates with linear
relations between the percentiles and $\log(Z)$. We obtain
$[(-37\pm20)\,\log(Z/0.0122)+108\pm 6]$ km~s$^{-1}$ for the median and $[(-36\pm35)\,\log(Z/0.0122)+190\pm 10]$ km~s$^{-1}$ for the $75^\mathrm{th}$ percentile.
The best-fitting value for the slope turns out to be negative as expected but the data points show substantial scatter around the regression line suggesting that other factors\footnote{The method used to measure the rotation velocities might also influence the results of the different papers. 
However, method comparisons
found small differences (e.g. \num{1} per cent in \citealt{Dufton2013} and below \num{10} per cent for nearly all stars in \citealt{Simon-Diaz2014}).}
beyond metallicity could play a role here. One of these is environment:
as it is well known that
stars in high-density clusters rotate faster than those in associations while field stars have even lower average rotation velocities \citep{Wolff2007}. A large fraction of undetected binaries could also influence the results. 

As a final step, we assume that ${\mathcal P}(\upsilon_\mathrm{ini})$ is well
approximated by a Weibull distribution. In order to determine the value of its two
free parameters as a function of $Z$, we derive the distribution of the projected
velocities following \citet{Gaige1993} and match its median and $75^\mathrm{th}$ percentile
to the fit we derived above.
The resulting ${\mathcal P}(\upsilon_\mathrm{ini})$
is then used to
weigh our stellar evolutionary models and compute
the feedback from a population of rotating stars.


\bsp	
\label{lastpage}
\end{document}